\documentclass[prd,twocolumn,aps,footinbib]{revtex4-2}
\usepackage{amsmath,amssymb,amsfonts,float,color,graphicx,graphics,latexsym,placeins,epsfig,multirow,bm,physics,bbold,enumitem,array,footmisc}

\usepackage{array}
\newcolumntype{P}[1]{>{\centering\arraybackslash}p{#1}}
\usepackage{footmisc}
\usepackage[compatibility=false]{caption}
\usepackage{subcaption}
\usepackage[font=small,labelfont=bf]{caption}
\usepackage{comment}
\usepackage{bbold}
\usepackage{enumitem}
\usepackage{braket}
\setdescription{leftmargin=8pt}
\usepackage{hyperref,url}
\hypersetup{
	colorlinks=true,       
	linkcolor=blue,          
	citecolor=blue,        
	urlcolor=blue           
}
\usepackage{natbib}
\setcitestyle{square, comma, numbers, sort&compress}
\usepackage{float}
\usepackage{amsmath,amssymb}

\usepackage{booktabs} 
\usepackage[compatibility=false]{caption}
\usepackage[font=small,labelfont=bf]{caption}
\usepackage{varioref}
\usepackage[active]{srcltx}
\usepackage{fancybox}
\usepackage[dvipsnames]{xcolor}

\def\a {\alpha}
\def\b {\beta}
\def\la {\lambda}
\def\om {\omega}
\def\th {\theta}
\def\a {\alpha}
\def\b {\beta}
\def\la {\lambda}
\def\om {\omega}
\def\th {\theta}

\newcommand{\Uu}{\boldsymbol{\mathcal{U}}}
\newcommand{\Vv}{\boldsymbol{\mathcal{V}}}
\newcommand{\Aa}{\boldsymbol{A}}
\newcommand{\Bb}{\boldsymbol{B}}
\newcommand{\Ll}{\boldsymbol{L}}
\newcommand{\Mm}{\boldsymbol{M}}
\newcommand{\Nn}{\boldsymbol{N}}
\newcommand{\Oo}{\boldsymbol{O}}

\newcommand{\C}{\mathcal{C}}
\newcommand{\B}{\mathcal{B}}
\newcommand{\A}{\mathcal{A}}
\newcommand{\D}{\mathcal{D}}

\newcommand{\ba}{\begin{align}}
\newcommand{\ea}{\end{align}}

\def\3nab{\tilde{\nabla}}

\def\be {\begin{equation}}
\def\ee {\end{equation}}
\def\ba {\begin{eqnarray}}
\def\ea {\end{eqnarray}}

\newcommand{\sfr}[2]
{{\textstyle\frac{#1}{#2}}}

\newcommand{\barray}{\begin{array}}
\newcommand{\earray}{\end{array}}

\newcommand{\bea}{\begin{eqnarray}}
\newcommand{\eea}{\end{eqnarray}}

\usepackage[dvipsnames]{xcolor}

\begin{document}
%
\title{\Large \bf Beyond Entanglement: Diagnosing quantum mediator dynamics in gravitationally mediated experiments}

\author{P. George Christopher }
\email{p.georgechristopher@iitb.ac.in}

\author{S. Shankaranarayanan}
\email{shanki@iitb.ac.in}

\affiliation{Department of Physics,  Indian Institute of Technology Bombay, Mumbai 400076, India}

\begin{abstract}
    No experimental test to date has provided conclusive evidence on the quantum nature of gravity. Recent proposals, such as the BMV experiment, suggest that generating entanglement could serve as a direct test. Motivated by these proposals, we study a system of three-harmonic oscillator system, with the mediator oscillator operating in two distinct parameter regimes: a heavy mediator regime and a light mediator regime. These regimes induce qualitatively different entanglement dynamics between the terminal oscillators. Crucially, distinguishing these regimes experimentally remains challenging when relying solely on entanglement measures. We
    demonstrate that the dynamical fidelity susceptibility offers a \emph{viable and sensitive probe} to contrast the regimes in practice. Our results provide testable signatures for optomechanical and trapped-ion platforms simulating gravitational interactions, and provide new avenues to characterize quantum-gravity-inspired systems beyond entanglement-based protocols.
\end{abstract}
\maketitle
\section{Introduction}
\label{sec:intro}
The Bose-Marletto-Vedral (BMV) proposal~\cite{Bose:2017nin,Marletto:2017kzi} has emerged as a promising avenue to experimentally probe the quantum nature of gravity. It proposes that two mesoscopic masses, each prepared in a spatial superposition, would become entangled via their gravitational interaction \emph{if and only} if gravity is quantum, a signature absent if mediated by a classical field~\cite{Oppenheim:2022xjr,Bose:2022uxe,Bose:2023gwh}.
While conceptually elegant, the BMV protocol faces dual challenges: Experimental challenges include maintaining coherence and detecting extremely weak gravity induced entanglement~\cite{Castro-Ruiz:2019nnl,Millen:2020brv,Westphal:2020okx,Carney:2021yfw,Fuchs:2023ajk}. Theoretical ambiguities persist: studies question whether entanglement is an unambiguous signature of quantum gravity, given that classical systems can mimic certain statistical correlations or even Bell-local correlations under certain conditions~\cite{Coradeschi:2021szx,Carney:2018ofe,Altamirano:2016fas,Ma:2021rve,Belenchia:2021rfb,Fragkos:2022tbm,Hanif:2023fto,Galley:2023byb,Martin-Martinez:2022uio,Pedernales:2023cwy,Etezad-Razavi:2023kgt}.
Motivated by these crucial interpretational challenges in gravitationally-mediated entanglement experiments, we develop and analyze a theoretical model of three harmonically coupled oscillators, that effectively mimic the BMV setup with generic quadratic couplings. Our investigation 
delineates what mediator-induced entanglement can \emph{genuinely reveal} about the fundamental nature of the mediating interaction itself.

    {In this work, we resolve \emph{this ambiguity} through a solvable three-coupled harmonic oscillator (HO) model, where oscillator $B$ acts as a tunable mediator. Our construction mirrors the essential assumptions of the BMV proposal: the initial global state is separable, and subsystems $A$ and $C$ couple only locally to the mediator $B$, never directly to each other. As a result, any entanglement observed between $A$ and $C$ can arise only through the mediator. This allows us to test, in a controlled setting, the central BMV idea that mediator-induced entanglement can act as an indicator of the mediator’s quantum nature. While our three-oscillator setup is not a literal model of gravity --- for instance, the mediator is a massive oscillator, not a massless graviton --- it serves as a powerful analogue framework. The mediator’s mass effectively sets the interaction range, enabling us to explore distinct dynamical regimes and draw physical insights into BMV-type scenarios.}


Within this framework, we explore two contrasting regimes: one where the mediator itself is highly squeezed, and another where the outer oscillators are predominantly squeezed. We first derive an analytical expression for the entanglement (quantified by Logarithmic Negativity ($\mathcal{E}_{\mathcal{N}}$) ~\cite{Plenio2005,Vidal:2002zz}) in the most symmetric weak-coupling scenario. For generic couplings and various initial states, we then employ numerical simulations to precisely map the entanglement patterns generated in these two extreme regimes. This allows us to identify the inherent difficulties in using entanglement as the sole resource to characterize the mediator. Beyond entanglement, we also analyze the behavior of quantum discord (\(\mathcal{D}\)) and introduce \emph{dynamical fidelity susceptibility} (\(\chi_F)\)~\cite{Chandran:2024utf} as a novel diagnostic tool.
We show that $\chi_F$ clearly discriminates between the two regimes. Finally, by prescribing parameter tuning (mediator mass, coupling strengths) and monitoring $\chi_F$, we provide practical guidance to avoid null-result pitfalls in BMV-type experiments.

\begin{figure}[h]
    \hspace*{-1.0cm}   \includegraphics[width=10cm]{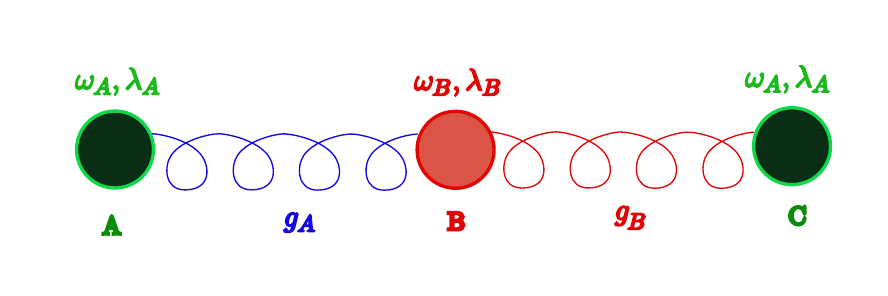}
    \caption{\label{fig1} Schematic diagram of the tripartite HO system. Oscillator B mediates interaction with A and C. Interactions include self-squeezing $(\lambda_A, \lambda_B)$ and inter-oscillator couplings $(g_A, g_B)$, like parametric squeezing and mode mixing.}
\end{figure}
\section{The model and set up}
\label{sec:setup}
Consider three 1-D quantum HOs, as depicted in Fig.~\eqref{fig1}. We treat the oscillators \textit{A} and \textit{C} as identical HOs with same shape, mass (\(m_{A}\)) and frequency (\(\om_{A}\)) and the oscillator \textit{B} acts as a local mediator connecting \textit{A} and \textit{C}, which do not interact directly. The HOs \textit{A} and \textit{C} and their interaction with the mediator \textit{B} is modelled via the Hamiltonian (in units of \(\hbar\))
\begin{eqnarray}
    \hat{H}  = \hat{H}^{(0)}_{A} + \hat{H}^{(0)}_{B} + \hat{H}^{(0)}_{C} + \hat{H}_{AB} + \hat{H}_{BC} \label{eq:1}
\end{eqnarray}
where the superscript ${(0)}$ denote the free Hamiltonians of the individual HOs along with self squeezing interaction strength ($\lambda_A, \lambda_B$), while the last two terms describe the interactions strengths $(g_A, g_B)$:
\begin{subequations}
    \label{eq:Hamiltonian}
    \begin{eqnarray}
        \hat{H}^{(0)}_{i} &=& \omega_{i}\left(a_{i}^{\dag} a_i + \mathbb{I}/2\right) + \lambda_{i}\left(a_i^{2} + a_i^{\dag 2}\right)/2 \, , \label{eq:2a} \\
        \hat{H}_{ij} &=&  g_{i}\left(a_i^{\dag} a_j + a_i a_j^{\dag}\right) +  g_{i}\left(a_i a_j + a_i^{\dag } a_j^{\dag}\right)\, ,  \label{eq:2d}
    \end{eqnarray}
\end{subequations}
where $i,j = \textit{A,B,C}$ and $a_i = (a,b,c)$ respectively for $i = A,B,C.$ The annihilation and creation operators
(\(a , b, c\)) and (\(a^{\dag} , b^{\dag}, c^{\dag}\)) satisfy the standard bosonic commutation relations.
The Hamiltonian incorporates three types of interactions: \textit{self-squeezing}, \textit{parametric squeezing}, and \textit{mode mixing}. The interaction between the mediator and the other HOs resembles those in the \emph{Caldeira-Leggett model}, coupling subsystems via terms like $x_A x_B$ or $x_B x_C$~\cite{Caldeira:1982iu}.
These inter-oscillator couplings decompose into two qualitatively distinct components: mode mixing, which conserves particle number (e.g., $a^\dagger b + ab^\dagger$), and parametric squeezing, which does not (e.g., $ab+a^\dagger b^\dagger$)~\cite{Rugar:1991ab}. {The parametric squeezing and mode-mixing terms in our Hamiltonian are reminiscent of the effective Newtonian interaction that would arise between two point masses at large separation. However, unlike a direct two-body coupling that would directly entangle $A$ and $C$, in our construction the interaction is realized locally via the mediator $B$, while still retaining the essential features of a Newtonian potential.} This decomposition facilitates non-trivial energy redistribution and building of correlations across subsystems.

The inclusion of self-squeezing terms in the above Hamiltonian, similar to those observed in optomechanical systems, is motivated by the non-linear nature of gravity, where even in the linear limit, phenomena like gravitational memory exist~\cite{Zeldovich:1974gvh,1987-ThorneBraginskii-Nature, Chakraborty:2024ars}. Thus, self-squeezing is crucial for model aiming to shed light on gravitational interactions. {Taken together, these ingredients ensure that the model retains the key structural features associated with gravitational mediation: interactions are strictly local, the coupling structure echoes that of a Newtonian potential, and nonlinear self-effects are incorporated through the squeezing terms. Crucially, this controlled setup allows us to probe in a precise way what the subsequent emergence of entanglement between A and C actually signals about the quantum character of the mediator.}
%
To study the system's dynamics, we employ the covariance matrix formalism~\cite{Eisert:2008ur} {(See Appendix~\eqref{sec:Sympformalism} for details.)} First, we express Hamiltonian \eqref{eq:1} as:
\begin{equation}
    \hat{H} = \mathbb{X}^{\dag}H\mathbb{X}/2~~{\rm where}~~
    H = \begin{pmatrix}
        \boldsymbol{U} & \boldsymbol{V} \\
        \boldsymbol{V} & \boldsymbol{U}
    \end{pmatrix} \, ,
    \label{eq:11}
\end{equation}
\(\hat{\mathbb{X}} = \left[a, b, c, a^{\dag}, b^{\dag}, c^{\dag}\right]^{T}\) is the column vector of mode operators. The submatrices \(\boldsymbol{U}\) and \(\boldsymbol{V}\) are given by
\begin{align}
    \boldsymbol{U} = \begin{pmatrix}
                         \om_{A} & g_{A}   & 0       \\
                         g_{A}   & \om_{B} & g_{B}   \\
                         0       & g_{B}   & \om_{A} \\
                     \end{pmatrix} \quad
    \boldsymbol{V} = \begin{pmatrix}
                         \lambda_{A} & g_{A}       & 0           \\
                         g_{A}       & \lambda_{B} & g_{B}       \\
                         0           & g_{B}       & \lambda_{A}
                     \end{pmatrix} \, .\label{eq:12}
\end{align}
The Hamiltonian \eqref{eq:1} is quadratic, guaranteeing that the time evolution preserves Gaussianity~\cite{adesso:2014}. Consequently, if the system is initially in a Gaussian state, it remains Gaussian throughout~\cite{Chandran:2022vrw}. We therefore restrict our analysis to Gaussian states ($\hat{\rho}_g$), which are fully characterized by their second moments of the quadrature operator vector \(\hat{\mathbb{X}}\) and are defined as:
\begin{align}
    \boldsymbol{\sigma}_{nm} = \langle \{\hat{\mathbb{X}}_n, \hat{\mathbb{X}}^{\dag}_m\} \rangle_{g}
    - 2 \langle \hat{\mathbb{X}}_n \rangle_{g} \langle \hat{\mathbb{X}}_m^{\dag} \rangle_{g} \label{eq:14}
\end{align}
where \(\langle \cdot \rangle_{g}\) denotes the expectation value of the operator with respect to state \(\hat{\rho}_{g}\) and \{\(\cdot \)\} is the anticommutator.

The time evolution of the moments is governed by the symplectic transformation \(S_H(t) = \exp(\Omega H t)\), where the canonical symplectic form for this complex-operator basis is \(\Omega = -i\,\text{diag}(1,1,1,-1,-1,-1)\). This transformation preserves the symplectic structure, \(S_H(t)\,\Omega\,S_H^T(t) = \Omega\), thereby ensuring that the commutation relations remain invariant. The second moments evolve as: $\boldsymbol{\sigma}(t) = S_H(t)\,\boldsymbol{\sigma}(0)\,S_H^\dagger(t)$.
To investigate the development of quantum correlations in this system, we analyze the time evolution of the covariance matrix \(\boldsymbol{\sigma}(t)\) for the Hamiltonian~\eqref{eq:1}. In realistic experimental scenarios --- such as the \emph{BMV-type setups} --- gravitationally mediated entanglement between two massive quantum systems (e.g., \textit{A} and \textit{C}) would be inferred from measurements performed over multiple experimental runs. Since experiments access only ensemble-averaged data, we evaluate \emph{long-time averages} of (\(\mathcal{E}_{\mathcal{N}}\)), computed over timescales much longer than the inverse normal-mode frequencies.
This approach qualitatively captures the  entanglement structure mediated by oscillator \textit{B}, reflecting what would be observed in ensemble measurements across many experimental runs. \\[1pt]

\section{Time evolution of the Hamiltonian}
\label{sec:Time-evolve}
The time evolution of the covariance matrix $\boldsymbol{\sigma}(t)$ is obtained from the system's normal mode frequencies $(k_1,k_2,k_3)$ and the Bogoliubov coefficients \(\boldsymbol{\alpha}\) and \(\boldsymbol{\beta}\), combined with the initial state's covariance matrix (see Appendix~\eqref{sec:3CHOsol} for details).
To analyze quantum correlations between subsystems \textit{A} and \textit{C}, we trace out the mediator \textit{B} from the time-evolved covariance matrix $\boldsymbol{\sigma}(t)$ to obtain the reduced state $\boldsymbol{\sigma}_{A-C}(t)$. The logarithmic negativity $\mathcal{E}_{\mathcal{N}}(t)$ and quantum discord \(\mathcal{D}(t)\) are then computed using standard techniques for Gaussian states~\cite{Vidal:2002zz}.

We proceed to analyze the system dynamics across a range of initial states and couplings, examining how quantum correlations depend on both system parameters and initial state preparation. This allows us to understand how the presence or absence of entanglement constrains the system and initial state parameters. As a starting point, we consider the most general separable Gaussian initial state: a product of single-mode squeezed states for each oscillator:
\begin{equation}
    \ket{\psi}_{0} = \ket{r_{1},\theta_{1}}\otimes\ket{r_{2},\theta_{2}}\otimes\ket{r_{3},\theta_{3}} , \label{eq:9}
\end{equation}
where each $\ket{r,\theta}$ is a single-mode squeezed vacuum state:
\begin{align}
    \ket{r,\theta} = S(\zeta)\ket{0}, \quad S(\zeta) = \exp{\left[ (\zeta^{*} \hat{a}^{2} - \zeta \hat{a}^{\dagger 2})/2\right]} \label{eq:42}
\end{align}
Here, $\zeta = r e^{i \theta}$ is the complex squeezing parameter, with $r$ ($\theta$) as the squeezing strength (angle)~\cite{Caves:1981}. While analytical solutions for $\mathcal{E}_{\mathcal{N}}(t)$ and $\mathcal{D}(t)$ are, in principle, possible for general Gaussian initial states, the complexity in asymmetric parameter regimes makes this approach cumbersome. Therefore, our primary method is numerical simulation to explore the full parameter space.

To gain physical insight, we obtain an \emph{analytical expression} for $\mathcal{E}_{\mathcal{N}}$ in the weak coupling limit under a highly simplified configuration. This specific scenario assumes an initial vacuum state  ($r_1 = r_2 = r_3 = 0, \theta_1 = \theta_2 = \theta_3 = 0$), frequency-degenerate oscillators ($\omega_A = \omega_B = \omega$), symmetric couplings to the mediator ($g_A = g_B = g$), and no local squeezing interactions ($\lambda_A = \lambda_B = 0$). In this regime, we consider the limit $\lambda_p \equiv \sqrt{2}g/\omega \ll 1$. We find that the long-time averaged logarithmic negativity $\langle \mathcal{E}_{\mathcal{N}} \rangle$ between subsystems \textit{A} and \textit{C} takes the form:
\begin{equation}
    \langle \mathcal{E}_{\mathcal{N}} \rangle \approx {\lambda_p}/{(\pi \ln 2)} =(\sqrt{2} g)/({\pi \ln 2 \omega})
\end{equation}
Interestingly, in the weak coupling regime, the average entanglement varies inversely with the mediator frequency $\omega$, implying that lower-frequency mediators facilitate stronger effective correlations. However, the apparent divergence as $\omega \to 0$ is unphysical because the weak coupling condition ($\lambda_p \ll 1$) requires $g$ to vanish at least proportionally to $\omega$. This behavior is nonetheless reminiscent of entanglement enhancement seen in bipartite systems featuring zero modes~\cite{Chandran_2019,Jain:2021ppx}. \\[1pt]
\begin{figure}[t]
    \centering
    \includegraphics[width=0.9\linewidth]{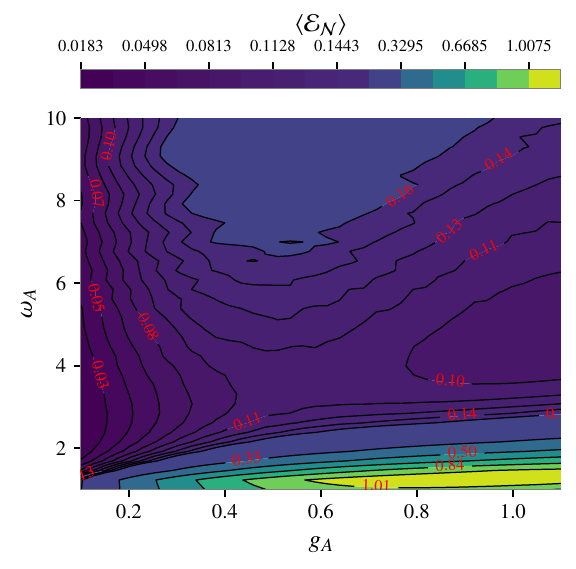}
    \includegraphics[width=0.9\linewidth]{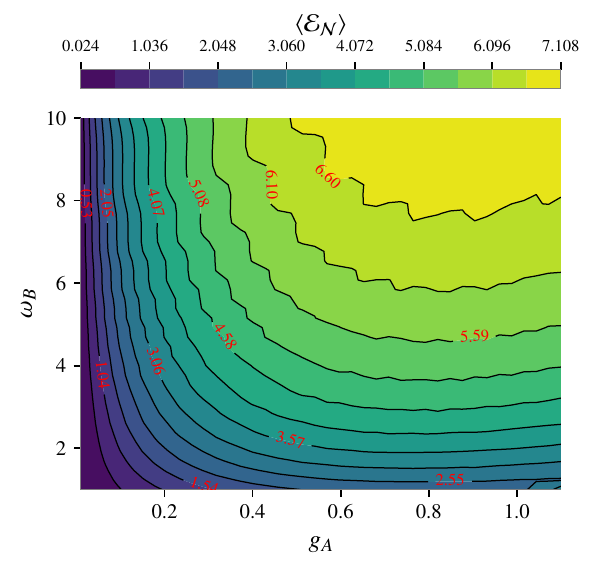}
    \caption{\label{fig:2}
    Contour plots of the long-time averaged logarithmic negativity
    $\langle \mathcal{E}_{\mathcal{N}} \rangle$ in the two mediator regimes.
        {\bf Top:} (HMR)
    {\bf Bottom:} (LMR).
    Parameter choices and initial squeezed states are specified in the text.
    }
\end{figure}

\section{Numerical Analysis and physical regimes}
%
To move beyond the analytically tractable weak-coupling regime, we numerically simulate the system's dynamics using the full Bogoliubov coefficients ($\boldsymbol{\alpha}, \boldsymbol{\beta}$) (See Table I Appendix~\eqref{sec:3CHOsol} for details). In what follows, we demonstrate our analysis on two distinct physical regimes starting from an initially squeezed state and provide physical arguments as to what is happening to the mediator in the two regimes
\begin{enumerate}[leftmargin=0.4cm]
    \item {\bf HMR} (Heavy Mediator Regime): This regime corresponds to the limit $\lambda_B \to \omega_B$, with $1 - \lambda_A/\omega_A$ remaining finite. In this limit, the mediator oscillator \textit{B} behaves as an effectively heavy, quasi-classical field mode. To understand its physical implications, consider the single-oscillator Hamiltonian with self-squeezing (cf. Eq.~\eqref{eq:2a}). When expressed in the position–momentum basis, after transformation, the Hamiltonian simplifies to:
          \begin{equation}
              \label{eq:HamiltonianB}
              \hat{H}_{0B} = {\hat{p}_B^2}/{(2\tilde{m}_B}) + \tilde{k}_B\, \hat{x}_B^2/2,
          \end{equation}
          Here, $\tilde{m}_B$ and $\tilde{k}_B$ are
          given by:
          \begin{equation}
              \tilde{m}_B = \frac{m_B}{1 - \lambda_B/\omega_B},
              \tilde{\omega}_B^2 = \omega_B^2 - \lambda_B^2,
              \tilde{k}_B = \tilde{m}_B\tilde{\omega}_B^2,
          \end{equation}
          where $m_B$ is the bare mass of the mediator. In the critical limiting case $\lambda_B \to \omega_B$, the effective mass $\tilde{m}_B$ diverges ($\tilde{m}_B \to \infty$) while the renormalized frequency vanishes ($\tilde{\omega}_B \to 0$). Crucially, their product $\tilde{k}_B = \tilde{m}_B\tilde{\omega}_B^2$ remains finite and equal to twice the bare spring constant $m_B \omega_B^2$. Physically, this causes the kinetic term to vanish in Hamiltonian \eqref{eq:HamiltonianB}, i.e.,
          \begin{equation}
              \hat{H}_{0B} \xrightarrow[\lambda_B \to \omega_B]{} \tilde{k}_B \, \hat{x}_B^2/2.
          \end{equation}
          This limit corresponds to an extreme self-interaction that effectively suppresses the canonical momentum $\hat{p}_B$, making $\hat{x}_B$ a conserved quantity. Hence, the oscillator no longer evolves dynamically but acts as a static background field fixed by its initial position $\hat{x}_B(0)$. This behavior is analogous to a cosmological constant~\cite{peebles:2002}, where vacuum energy is given by the constant scalar field potential, here represented by a fixed energy density parameterized by $\hat{x}_B(0)^2$.

          Thus, in HMR, the mediator loses its ability to coherently store or transmit quantum information. Its Hilbert space remains intact, but its dynamical role is effectively frozen: the canonical momentum $\hat{p}_B$ vanishes from the Hamiltonian, and the position $\hat{x}_B$ becomes a conserved quantity. This reduction mirrors constrained dynamics in gauge theories and systems with Hamiltonian constraints, where certain degrees of freedom become dynamically redundant or decouple from the evolution~\cite{henneaux1992quantization,rovelli2004quantum}.

    \item {\bf LMR} (Light Mediator Regime): This regime corresponds to $\lambda_A \to \omega_A$, while $1 - \lambda_B/\omega_B$ remains finite. Here, mediator \textit{B} is not significantly self-squeezed, remaining highly quantum and susceptible to fluctuations.  Fig.~\eqref{fig:2} (bottom) contains the plot of $\langle \mathcal{E}_{\mathcal{N}} \rangle$ as a function of $\omega_B$ and $g_B$ by fixing $\omega_A$, $\lambda_A$, $\lambda_B$, and $g_A$.
\end{enumerate}

\begin{figure}[t]
    \centering
    \includegraphics[width=0.49\linewidth]{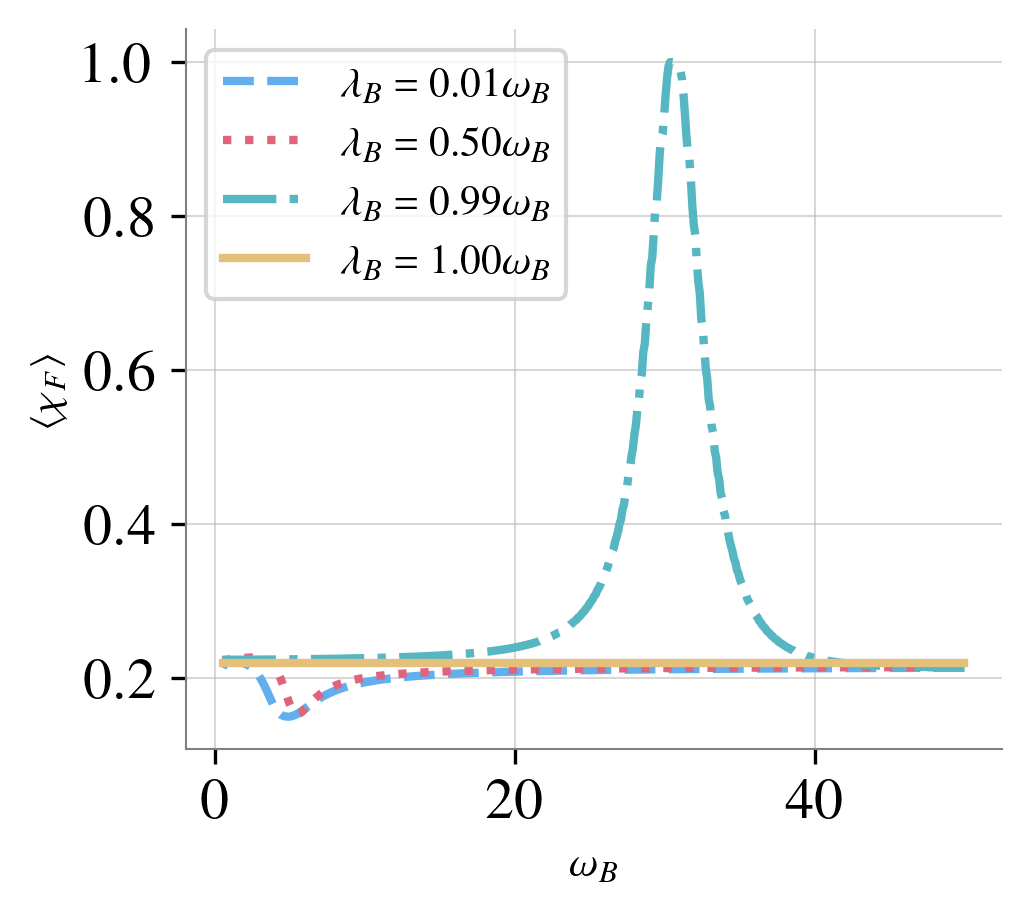}
    \includegraphics[width=0.49\linewidth]{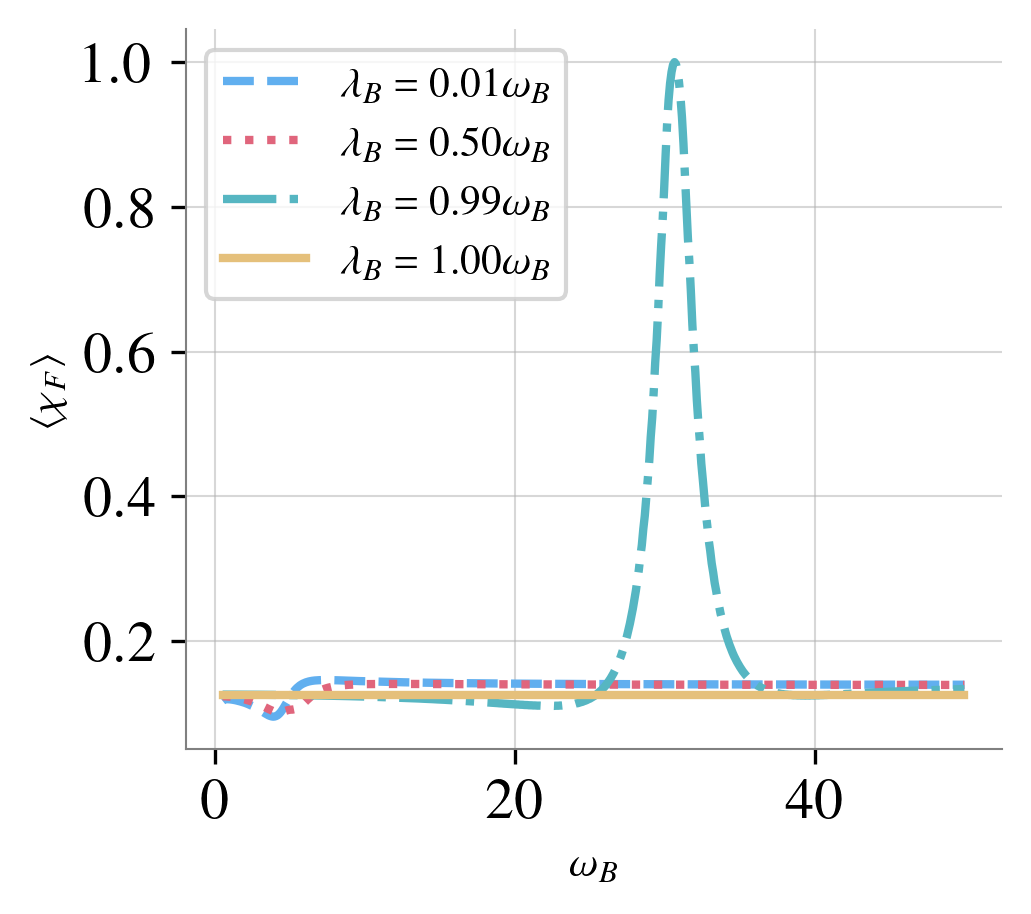}
    \caption{\label{fig:3}
        Normalized time-averaged $\langle \chi_F \rangle$ vs $\omega_B$ for two different initial states (vacuum: left, squeezed: right) with $r_1 = r_3 = 0.5$, $r_2 = 0.2$, \(\th_1 = \th_2 = \th_3 = 0\).
        Fixed parameters: $\omega_A = 5$, $\lambda_A = 0.5\,\omega_A$, $g_A = g_B = 0.6$.}
\end{figure}

To visualize how entanglement between subsystems \textit{A} and \textit{C} depend on system parameters in both mediator regimes, we plot contour maps of the long-time averaged $\langle \mathcal{E}_{\mathcal{N}} \rangle$. For the HMR [Fig.~\ref{fig:2} (top)], we take $(\omega_B,\lambda_B,\lambda_A,g_B)=(5.0,4.9,10^{-4},1.0)$ with initial state $r=(0.1,1.0,0.1)$ and $\theta=(-\pi/3,\pi/4,\pi/6)$, and plot $\langle \mathcal{E}_{\mathcal{N}} \rangle$ as a function of $(\omega_A,g_A)$.
For the LMR [Fig.~\ref{fig:2} (bottom)], we set $(\omega_A,\lambda_A,\lambda_B,g_B)=(5.0,4.999,\omega_B/1.5,1.4)$ with $r=(0.5,2.0,0.5)$ and $\theta=(0,0,0)$ and plot $\langle \mathcal{E}_{\mathcal{N}} \rangle$ as a function of $(\omega_B,g_A)$. From numerical simulations, we also confirm that similar behavior is obtained even for the case where the system starts out in a vacuum state initially. Further $\langle {\cal D} \rangle$ shows similar behaviour as that of entanglement (see Figs.~\ref{fig:4} and \ref{fig:5}, Appendix~\eqref{sec:Plots})
. The contour plots of $\langle\mathcal{E}_{\mathcal{N}}\rangle$ in both the {\bf LMR} and {\bf HMR} regimes indicate that entanglement between the two subsystems can be mediated via oscillator $B$, though through qualitatively distinct mechanisms.
To further differentiate these regimes and characterize the mediator's behavior, we compute the \emph{time-averaged dynamical fidelity susceptibility} $\langle \chi_F \rangle$~\cite{Chandran:2024utf}.{ In order to capture the mediator’s effect on the two subsystems, this quantity is evaluated for the reduced state of $A$–$C$, described by the covariance matrix $\sigma_{AC}(t)$. The dynamical fidelity $\mathcal{F}(t)$ for mixed states is given by
\begin{align}
    \mathcal{F}_{\text{mixed}} = \mathrm{Tr}\!\left[\sqrt{\rho^{1/2}(t)\,\rho(t + \delta t)\,\rho^{1/2}(t)}\right].
\end{align}
Expanding $\mathcal{F}(t)$ for small $\delta t$ yields
\begin{align}
    \!\!\!\!
    \mathcal{F}(t,\delta t) \sim 1 - \chi_F(t)\,(\delta t)^2,
    ~ & ~
    \chi_F(t) \equiv -\tfrac{1}{2}\,\left.\frac{d^2 \mathcal{F}}{d(\delta t)^2}\right|_{\delta t=0}
    \label{eq:DFS}
\end{align}
where $\chi_F(t)$ is the \textit{dynamical fidelity susceptibility (DFS)}.
To compute $\chi_F(t_0)$, we take $\sigma_{AC}(t_0)$ as the reference state and evaluate its overlap with $\sigma_{AC}(t_0 \pm \delta t)$
via the Gaussian fidelity expression provided in Appendix~\eqref{sec:QMeasures}, from which $\chi_F$ is extracted.}
We then evaluate the time-averaged $\langle \chi_F \rangle$ as a function of the mediator frequency $\omega_B$, as shown in Fig.~\eqref{fig:3}. Our results reveal distinct variations in $\langle \chi_F \rangle$ depending on the system's proximity to the HMR.
For parameter values ($\lambda_B$) {away from the HMR},
the susceptibility exhibits a characteristic inflection point in the vicinity of $\omega_B \approx \omega_A$ --- a feature absent in the near-HMR and exact HMR regimes. In this regime, $\langle \chi_F \rangle$ shows heightened sensitivity only within a narrow region around $\omega_B \approx \omega_A$, saturating to different baseline values elsewhere. By contrast, {in the proximity of the HMR limit} ($\lambda_B \approx \omega_A$), the response of $\langle \chi_F \rangle$ sharpens into a prominent, resonance-like peak. Beyond this peak, the susceptibility rapidly falls off and saturates to the flat HMR value corresponding to the critical point $\lambda_B = \omega_B$.
The nature of this sensitivity fundamentally differs across regimes: near the HMR, $\langle \chi_F \rangle$ is highly sensitive to variations in $\omega_B$ {away from} $\omega_A$, reflecting a broad change in the mediator's dynamics. Conversely, far from the HMR, the susceptibility is enhanced {only in the immediate vicinity} of $\omega_B \approx \omega_A$, indicating a localized sensitivity. These distinct trends powerfully demonstrate that $\langle \chi_F \rangle$ serves as a robust diagnostic of the mediator's operational regime. Plots for the case $\lambda_A = 0.01\,\omega_A$ are provided in Appendix~\eqref{sec:DSF}.

In our model, the mediator's intrinsic character is determined by the parameters $\omega_B$ and $\lambda_B$. Varying $\omega_B$ effectively simulates different mediator regimes. Within BMV-type experimental scenarios, where the mediator is gravitational, tuning effective parameters analogous to $\omega_B$ could correspond to probing different gravitational regimes, such as variations in the local spacetime curvature $\mathcal{R}$. The quantity $\sqrt{\mathcal{R}}$ then sets a natural frequency scale associated with the hypothesized gravitational mediator. In such a setting, system parameters such as $\omega_A$ can be chosen to be comparable to $\sqrt{\mathcal{R}}$, and varied around this scale to explore the behavior of the time-averaged dynamical fidelity susceptibility. Measuring and comparing $\langle \chi_F \rangle$ across such regimes could provide operational insight into whether the mediator exhibits fundamentally quantum or effectively classical behavior—thereby offering a novel experimental pathway to characterize gravitationally-mediated interactions.

\section{Results and Implications}
%
{Our numerical analysis, presented in Fig.~\eqref{fig:2}, reveals that in both the HMR and the LMR, the mediator can facilitate the establishment of {quantum correlations between \textit{A} and \textit{C}, but with different structural and spectral features in the two regimes.}{ The LMR consistently displays a smoother and more continuous entanglement landscape, with nonzero quantum correlations broadly distributed across wide parameter ranges [Fig.~\ref{fig:2} (bottom), Fig.~\ref{fig:4}(c) in Appendix~\eqref{sec:Plots}]. This indicates a robust ability of the dynamically active, ``light'' mediator $B$ to facilitate entanglement between $A$ and $C$. When the mediator can fluctuate significantly, it readily mediates quantum correlations. While at isolated instants the bipartite entanglement between $A$ and $C$ may transiently vanish due to the global state resembling a continuous variable-GHZ state, our analysis captures the time-averaged behavior over dynamical cycles, which always yields nonzero correlations in this regime.}

{In contrast, in the HMR, bipartite entanglement emerges only within a sharply localized patches of the $(\omega_A, g_A)$ parameter space [Fig.~\ref{fig:2} (top), Fig.~\ref{fig:4}(a) in Appendix~\eqref{sec:Plots}], while outside this region the time-averaged logarithmic negativity is strongly suppressed or vanishes in the exact limit (\(\lambda_B = \omega_B\))(See Appendix~\eqref{sec:DSF} for details). This arises from fine-tuned parameter choices that bring the system close to resonance, enabling transient enhancement
even at finite coupling strengths $g_A$ and $g_B$. By contrast, in the LMR such suppression only occurs for very small coupling strengths.

This behaviour is reminiscent of \emph{quantum many-body scars}, in which non-generic eigenstates maintain unexpectedly low entanglement and resist thermalization~\cite{Moudgalya:2021xlu,Deutsch:2018ulr}. Here, we observe an analogous form of scars, not in a many-body Hilbert space but within the space of Gaussian states of this continuous-variable system. {Further, the suppression of entanglement observed in the HMR may superficially resemble entanglement sudden death, but in our integrable Gaussian setting complete disappearance of correlations occurs only in the strict zero-mode limit of the mediator (for details, see Appendix~\eqref{sec:HMR})}.

{This asymmetry arises from the eigenmode structure: in HMR ($\lambda_B \to \omega_B$), the mediator's near degenerate eigenmodes ($k_2 \to 0$, $k_1 \to k_3$)
induce entanglement suppression, demonstrating that quantum systems \emph{can appear} classical under dynamical constraints, while in LMR ($\lambda_A \to \omega_A$), the two eigenmodes vanish simultaneously ($k_1, k_2 \to 0$), with the third remaining finite.} This suggests persistent entanglement arises from the mediator’s active fluctuations confirming its quantum role. These contrasting signatures highlight an inherent asymmetry in the mediator’s dynamical role across the two regimes and may have implications for relativistic systems~\cite{Fragkos:2022tbm,Martin-Martinez:2022uio}.

These results carry significant implications for experimental tests of quantum gravity, such as BMV-type scenarios where two massive systems interact via a presumed quantum mediator. Such scenarios are being explored with optomechanical oscillators and trapped ions~\cite{Teufel2011, Chan2011, Abbott2009, Krisnanda2019}.{\;Our findings demonstrate that a mediator treated as quantum can establish entanglement while existing in two physically distinct regimes: in the LMR it actively establishes entanglement, while in the HMR it becomes dynamically constrained and correlations vanish at ($\lambda_B = \omega_B$). Thus, Entanglement-based witnesses alone may therefore be insufficient, as they cannot distinguish these regimes by themselves. Thus, absence of entanglement in BMV-type tests may indicate HMR-like freezing \emph{not classicality}.}

{When extended to gravity, our results show that a quantum mediator can exhibit signatures that appear classical, namely suppressed entanglement and frozen fluctuations. The dynamical fidelity susceptibility ($\langle \chi_F \rangle$) distinguishes these regimes, revealing when the mediator is dynamically constrained yet still quantum. Such constrained behavior may carry deeper meaning in quantum gravity, where self-interactions can localize information in the field, reminiscent of gravitational memory effects. Consequently, expected quantum signatures may be masked under realistic conditions, underscoring the need for probes beyond entanglement.}

{For BMV-type experiments, this means the mediator must operate in a regime where its quantum dynamics remain active—a nontrivial requirement given our limited knowledge of its internal structure. We show that tracking ($\langle \chi_F \rangle$) against system parameters, such as oscillator frequencies or effective mass, provides a direct handle on the predicted HMR–LMR transition. Thus, the BMV paradigm must be broadened: entanglement generation is sufficient but not necessary to establish a quantum mediator. Revised criteria should combine entanglement with complementary probes such as mediator tomography~\cite{Vidal:2025ona} and fidelity susceptibility.}

\begin{acknowledgments}
    The authors are grateful to S. M. Chandran, Himadri Dhar, K. Hari and Y. Ibrahim for their valuable discussions and feedback on the earlier draft. The work is supported by SERB-CRG/2022/002348.  The MHRD fellowship at IIT Bombay financially supports PGC.
\end{acknowledgments}

\appendix
\section{Symplectic formalism}
\label{sec:Sympformalism}

This section provides a brief overview of the covariance matrix and symplectic formalism, which is particularly well-suited for systems governed by Hamiltonian that are quadratic in the canonical quadrature operators (\(\hat{q}\) and \(\hat{p}\)), or equivalently in the Bosonic creation and annihilation operators (\(\hat{a}\) and \(\hat{a}^{\dagger}\)). Instead of working in the phase-space picture via wavefunctions or characteristic functions of type ``\(s\)'' (denoted \(\chi_s\)), a more elegant and computationally efficient approach involves the use of the covariance matrix formalism, especially for Gaussian states. For Gaussian states, knowledge of the first and second moments is sufficient to fully reconstruct the state, as all higher-order moments can be expressed in terms of them due to Wick’s theorem~\cite{Weedbrook:2012}.

We begin by defining a linear vector space (LVS) by collecting the annihilation and creation operators into a column vector of the form:
\[\hat{\mathbb{X}} = \left[\hat{a}_{1}, \hat{a}_{2},\ldots,\hat{a}_n,\hat{a}^{\dag}_{1},\hat{a}^{\dag}_{2},\ldots, \hat{a}^{\dag}_n\right]^{T}\]
of length ``\(2n\)'', where ``\(n\)'' represents the number of oscillator modes in the system. The canonical commutation relations between the various mode operators can now be recast as
\begin{align}
    \left[\hat{\mathbb{X}}_{i},\hat{\mathbb{X}}^{\dag}_{j}\right] = i \Omega_{ij} \label{eq:A1}
\end{align}
where, \(\Omega \) is called the symplectic form given by,
\begin{align}
    \Omega = -i \begin{pmatrix}
                    \mathbb{I}_{n} & \mathbb{O}_{n}  \\
                    \mathbb{O}_{n} & -\mathbb{I}_{n}
                \end{pmatrix} \label{eq:A2}
\end{align}
In the traditional density matrix formalism the time evolution of the system is given by a unitary operator \(\hat{U}(t)\) such that \(\rho_{0} \rightarrow \rho(t) = \hat{U}(t)\rho_{0}\hat{U}^{\dag}(t)\), which in the covariance matrix formalism corresponds to a symplectic transformation of the first and second order moments. The first order moments transform as
\(\langle\hat{\mathbb{X}}_{0}\rangle \rightarrow \langle\hat{\mathbb{X}}_{0}\rangle(t) = S_{H}(t)\langle\hat{\mathbb{X}}_{0}\rangle \) while the second order moments in the form of the covariance matrix transform as~\cite{Simon:1988}:
\[\sigma_{0} \rightarrow \sigma (t) = S_{H}(t) \sigma_{0} S_{H}^{T}(t) \, , \]
where the covariance matrix is defined as,
\begin{align}
    \sigma_{ij} = \langle \{\hat{\mathbb{X}}_{i},\hat{\mathbb{X}}_{j}^{\dag}\} \rangle - 2 \langle \hat{\mathbb{X}}_{i}\rangle \langle\hat{\mathbb{X}}^{\dag}_{j} \rangle \label{eq:A3}
\end{align}
For time independent Hammitonian's the symplectic evolution matrix is given by \(S_{H}(t) = \exp[\Omega H t]\) (\(\hbar = 1\)). Where \(H\) represents the Hamiltonian quadratic form, which appears when we write the total Hamiltonian in terms of the vector \(\hat{\mathbb{X}}\) as \(\hat{H} = \frac{1}{2}\hat{\mathbb{X}^{\dag}}H\hat{\mathbb{X}}\). It is in general a hermitian matrix having the following structure,
\begin{align}
    H = \begin{pmatrix}
            \boldsymbol{U}     & \boldsymbol{V}                   \\
            \boldsymbol{V}^{*} & \boldsymbol{U}^{*} \label{eq:A4}
        \end{pmatrix}
\end{align}
with \(U = U^{\dag}\) and \(V = V^{T}\) due to hermicity of the Hamiltonian operator \(\hat{H}\). This specific structure of \(H\) arises due to the basis that we have chosen for our LVS, had we chosen a real basis for the column vector \(\hat{\mathbb{X}_r} = [\hat{x}_{1}, \hat{x}_{2},\ldots,\hat{x}_{n},\hat{p}_{1},\hat{p}_{2}, \ldots, \hat{p}_{n}]^{T}\), then it is related to the column vector \(\hat{\mathbb{X}}\) by
\begin{equation}
    \hat{\mathbb{X}} = \mathbb{U}\;\hat{\mathbb{X}_r}\label{eq:A5}
\end{equation}
Where the basis changing matrix \(\mathbb{U}\) elegantly reads
\begin{equation}
    \mathbb{U} = \frac{1}{\sqrt{2}}\begin{pmatrix}
        \mathbb{I}_n & i\mathbb{I}_n  \\
        \mathbb{I}_n & -i\mathbb{I}_n
    \end{pmatrix}\label{eq:A6}
\end{equation}
Thus, any matrix \(\mathbb{A}_r\) in the real basis will be written as,
\begin{equation}
    A = \mathbb{U}\;\mathbb{A}_r\;\mathbb{U}^{\dag}\label{eq:A7}
\end{equation}
in the complex basis. Thus if we write \(\mathbb{A}_r\) in the form of block matrices,
\begin{equation}
    \mathbb{A}_r = \begin{pmatrix}
        \boldsymbol{L} & \boldsymbol{M} \\
        \boldsymbol{N} & \boldsymbol{O}
    \end{pmatrix}\label{eq:A8}
\end{equation}
The same matrix in the transformed basis takes the form,
\begin{equation}
    A = \begin{pmatrix}
        (\Ll + \Oo) + i(\Nn - \Mm) & (\Ll - \Oo) + i(\Nn + \Mm) \\
        (\Ll - \Oo) - i(\Nn - \Mm) & (\Ll + \Oo) - i(\Nn - \Mm)
    \end{pmatrix} \nonumber
\end{equation}
Thus, any matrix in the complex basis would have the same structure as that of Eq.~\eqref{eq:A4}. For the case where \(U\) and \(V\) are real matrices, \(H\) becomes a real symmetric positive definite (the antisymmetric component of \(H\) only contributes as a constant to the Hamiltonian and hence ignored ~\cite{Serafini:2017rrn}) \(2n \times 2n\) matrix.  The symplectic time evolution matrix \(S_{H}\) satisfies \(S_{H}\Omega S_{H}^{\dag} = S_{H}^{\dag} \Omega S_{H} = \Omega \), and in general can be written as,
\begin{align}
    S_{H}(t) = \begin{pmatrix}
                   \boldsymbol{A}(t)     & \boldsymbol{B}(t)     \\
                   \boldsymbol{B}^{*}(t) & \boldsymbol{A}^{*}(t)
               \end{pmatrix}\label{eq:A9}
\end{align}
Now according to Williamson's theorem, any positive definite symmetric \(2n \times 2n\) matrix can be diagonalised via a symplectic transformation and put in the Williamson normal form. Applying this theorem to \(H\), we obtain
\begin{align}
    H = S_0^{\dag}\boldsymbol{K}S_{0} \label{eq:A10}
\end{align}
where \(\boldsymbol{K}\) is a diagonal matrix of the form diag(\(k_{1},k_{2},\ldots k_{n}, k_{1},k_{2},\ldots,k_{n}\)) (\(k_{i} \in \mathbb{R}^{+}\) for \(i = 1,2,\ldots,n\)). The diagonal entries are called the symplectic eigenvalues of \(H\), which are eigenvalues of the matrix \(i\Omega H\). The general structure of the symplectic matrix \(S_0\) is again of the same form as~\eqref{eq:A4} and can be written as,
\begin{align}
    S_{0} = \begin{pmatrix}
                \boldsymbol{\alpha}    & \boldsymbol{\beta}                     \\
                \boldsymbol{\beta}^{*} & \boldsymbol{\alpha}^{*} \label{eq:A11}
            \end{pmatrix}
\end{align}
Where the matrices \(\boldsymbol{\alpha}\) and \(\boldsymbol{\beta}\) are ``\(n \times n\)''matrices called the Bogoliubov matrices. The symplectic condition \(S_{0}\Omega S_{0}^{\dag} = S_{0}^{\dag}\Omega S_{0} = \Omega \) imposes further conditions on the Bogoliubov matrices called the Bogoliubov identities given as,
\begin{subequations}
    \label{eq:A12}
    \begin{eqnarray}
        \boldsymbol{\alpha} \boldsymbol{\alpha}^{\dag} - \boldsymbol{\beta} \boldsymbol{\beta}^{\dag} = \mathbb{I}_{2n} \label{eq:A12a} \\
        \boldsymbol{\alpha} \boldsymbol{\beta}^{\dag} - \boldsymbol{\beta} \boldsymbol{\alpha}^{\dag} = \mathbb{O}_{2n} \label{eq:A13b}
    \end{eqnarray}
\end{subequations}
Using Eq.~\eqref{eq:A10}, we can simplify the symplectic time evolution matrix \(S_{H}(t)\) as follows
\begin{align}
    S_{H}(t) & = \exp[\Omega H t]                                           \nonumber    \\
             & = \exp[\Omega S_{0}^{\dag}\boldsymbol{K}S_{0} t] \nonumber                \\
             & = \exp[S_{0}^{-1}S_{0}\Omega S_{0}^{\dag}\boldsymbol{K}S_{0} t] \nonumber \\
             & = \exp[S_{0}^{-1}\Omega\boldsymbol{K}S_{0} t]          \nonumber          \\
             & = S_{0}^{-1}\exp[\Omega\boldsymbol{K} t]S_{0} \label{eq:A13}
\end{align}
However the symplectic form \(\Omega \) is a diagonal matrix, hence using \(S_{0}\Omega S_{0}^{\dag} = \Omega \), we can rewrite Eq.~\eqref{eq:A13} as
\begin{align}
    S_{H}(t) & = -\Omega S_{0}^{\dag} \Omega \exp[\Omega \boldsymbol{K} t] S_{0} \label{eq:A14}
\end{align}
Now if we introduce \(\tilde{k} = \) diag(\(k_{1},k_{2},\ldots,k_{n}\)) and expand the right hand side of Eq.~\eqref{eq:A14} we obtain the equations,
\begin{subequations}
    \label{eq:A15}
    \begin{eqnarray}
        \boldsymbol{A}(t) & = \boldsymbol{\alpha}^{\dag} e^{-i \tilde{k} t}\boldsymbol{\alpha} - \boldsymbol{\beta}^{T} e^{i \tilde{k} t} \boldsymbol{\beta}^{*} \label{eq:A15a}  \\
        \boldsymbol{B}(t) & = \boldsymbol{\alpha}^{\dag} e^{-i \tilde{k} t} \boldsymbol{\beta} - \boldsymbol{\beta}^{T} e^{i \tilde{k} t} \boldsymbol{\alpha}^{*} \label{eq:A15b}
    \end{eqnarray}
\end{subequations}
With \(\boldsymbol{A}^{T}(t) = \boldsymbol{A}(t) \) and \(\boldsymbol{B}(t) = -\boldsymbol{B}^{\dag}(t)\). Further using Eq.~\eqref{eq:A4} and Eq.~\eqref{eq:A10} we get the following equations,
\begin{subequations}
    \label{eq:A16}
    \begin{eqnarray}
        \boldsymbol{U} & = \boldsymbol{\alpha}^{\dag} \tilde{k} \boldsymbol{\alpha} + \boldsymbol{\beta}^{T} \tilde{k} \boldsymbol{\beta}^{*} \label{eq:A16a} \\
        \boldsymbol{V} & = \boldsymbol{\alpha}^{\dag} \tilde{k} \boldsymbol{\beta} + \boldsymbol{\beta}^{T} \tilde{k} \boldsymbol{\alpha}^{*} \label{eq:A16b}
    \end{eqnarray}
\end{subequations}

Thus the entire system consists of solving the set of Eqs.~\eqref{eq:A16} along with the constraint Eqs.~\eqref{eq:A12}, which gives the explicit expressions for \(\boldsymbol{\alpha},\;\boldsymbol{\beta}\) and \(\tilde{k}\). Once we obtain these quantities we can use Eqs.~\eqref{eq:A15} to obtain \(S_{H}(t)\) and then find \(\langle\hat{\mathbb{X}_0}\rangle(t)\) and \(\sigma(t)\). If at time ``\(t\)'', the covariance matrix is given as ,
\begin{align}
    \boldsymbol{\sigma}(t) = \begin{pmatrix}
                                 \boldsymbol{\mathcal{U}(t)}     & \boldsymbol{\mathcal{V}(t)}     \\
                                 \boldsymbol{\mathcal{V}^{*}(t)} & \boldsymbol{\mathcal{U}^{*}(t)}
                             \end{pmatrix} \label{eq:A17}
\end{align}
The explicit form for \(\boldsymbol{\mathcal{U}}(t)\) and \(\boldsymbol{\mathcal{V}}(t)\) can be obtained using the explicit form of \(S_{H}(t)\) from Eq.~\eqref{eq:A9}, which gives us,
\begin{equation}
    \begin{aligned}
        \Uu(t) & = \Aa(t)\Uu_{0}\Aa^{*}(t) + \Bb(t)\Vv^{*}_{0}\Aa^{*}(t)       \\
               & \quad - \Bb(t)\Uu^{*}_{0}\Bb(t) - \Aa(t)\Vv_{0}\Bb(t)         \\
        \Vv(t) & = \Aa(t)\Vv_{0}\Aa(t) + \Bb(t)\Uu^{*}_{0}\Aa(t)               \\
               & \quad - \Aa(t)\Uu_{0}\Bb^{*}(t) - \Bb(t)\Vv^{*}_{0}\Bb^{*}(t)
    \end{aligned}
    \label{eq:A18}
\end{equation}
\section{Solution to the general three Harmonic oscillator model}\label{sec:3CHOsol}
In this section, we present the explicit solution for the dynamics of the system discussed in main text. The Hamiltonian quadratic form (\(H\)) of the system is composed of purely real parameters, as a result the entire covariance formalism can be reformulataed assuming all the matrices to be real. Thus the set of Eqs.~\eqref{eq:A12} and Eqs.~\eqref{eq:A12} and \eqref{eq:A16} reduce to
\begin{subequations}
    \label{eq:A19}
    \begin{eqnarray}
        \mathbb{I}_{2n} & = \boldsymbol{\alpha} \boldsymbol{\alpha}^{T} - \boldsymbol{\beta} \boldsymbol{\beta}^{T}  \label{eq:A19a}                    \\
        \mathbb{O}_{2n} & = \boldsymbol{\alpha} \boldsymbol{\beta}^{T} - \boldsymbol{\beta} \boldsymbol{\alpha}^{T}  \label{eq:A19b}                    \\
        \boldsymbol{U}  & = \boldsymbol{\alpha}^{T} \tilde{k} \boldsymbol{\alpha} + \boldsymbol{\beta}^{T} \tilde{k} \boldsymbol{\beta} \label{eq:A19c}
        \\
        \boldsymbol{V}  & = \boldsymbol{\alpha}^{T} \tilde{k} \boldsymbol{\beta} + \boldsymbol{\beta}^{T} \tilde{k} \boldsymbol{\alpha} \label{eq:A19d}
    \end{eqnarray}
\end{subequations}
where \(\boldsymbol{U}\) and \(\boldsymbol{V}\) are given by Eq.~\eqref{eq:12}. To check whether the  system of equations at hand are consistent, we can count the number of independent unknowns and independent equations in them. Due to symmetry of the matrices \(\boldsymbol{U}\) and \(\boldsymbol{V}\), Eqs.~\eqref{eq:A19c} - \eqref{eq:A19d} constitute 12 independent equations in 21 unknowns. However, the 21 unknowns are all not independent and are constrained by Eqs.~\eqref{eq:A19a} -- \eqref{eq:A19b} which are 9 in number. Thus, we have (21 - 9) = 12 unknowns and 12 independent equations.  \\
We can recast these set of equations in a much more elegant way by introducing linear combinations of the Bogoliubov matrices as \(\Delta_{+} = \boldsymbol{\alpha} + \boldsymbol{\beta}\) and \(\Delta_{-} = \boldsymbol{\alpha} - \boldsymbol{\beta}\), then the Bogoliubov identies become a single matrix equation as \(\Delta_{+}\Delta^{T}_{-} = \Delta_{-}\Delta^{T}_{+} = \mathbb{I}_{n}\) (where \(n = 3\) for this case), we can do further manipulation to write the equations for \(\boldsymbol{U}\) and \(\boldsymbol{V}\) in a better way by again considering linear combination of them as \(\boldsymbol{U}_{+} = \boldsymbol{U} + \boldsymbol{V}\) and \(\boldsymbol{U}_{-} = \boldsymbol{U} - \boldsymbol{V}\) and the corresponding equations become,
\begin{subequations}
    \label{eq:A20}
    \begin{eqnarray}
        \boldsymbol{U}_{+} = \Delta^{T}_{+} \tilde{k} \Delta_{+} \label{eq:A20a} \\
        \boldsymbol{U}_{-} = \Delta^{T}_{-} \tilde{k} \Delta_{-} \label{eq:A20b}
    \end{eqnarray}
\end{subequations}
On utilizing the Bogoliubov identities we obtain much elegant looking equations
\begin{subequations}
    \label{eq:A21}
    \begin{eqnarray}
        \Delta_{-}\boldsymbol{U}_{+} = \tilde{k} \Delta_{+} \label{eq:A21a} \\
        \Delta_{+}\boldsymbol{U}_{-} = \tilde{k} \Delta_{-} \label{eq:A21b}
    \end{eqnarray}
\end{subequations}
For the given choice of matrices \(\boldsymbol{U}\) and \(\boldsymbol{V}\), we can explicitly find our the symplectic eigenvalues of \(H\) by writing the characterisitic polynomial of the matrix \(i\Omega H\) under the assumptions \(\omega_{A} > \lambda_{A}, \omega_{B} > \lambda_{B}\), which gives
\begin{eqnarray}
    f(\lambda) &=& (\lambda^{2} + \lambda_{A}^{2} - \omega_{A}^{2})(\lambda^{4} + R(\omega_{A}-\lambda_{A})(\omega_{B}-\lambda_{B})  \nonumber \\
    &&+\lambda^2(\lambda_{A}^2+\lambda_{B}^2 - \omega_{A}^2 - \omega_{B}^2)) \\
    R &=& (\omega_{A}+\lambda_{A})(\omega_{B}+\lambda_{B}) - 4(g_{A}^{2}+g_{B}^2) \nonumber\label{eq:A22}
\end{eqnarray}
Solution to \(f(\lambda) = 0\) gives the symplectic eigenvalues,
\begin{align}
    k_{1} & = \sqrt{\omega_{A}^{2} - \lambda_{A}^{2}} \label{eq:A23}                                                                                          \\
    k_{2} & = \frac{1}{\sqrt{2}}\sqrt{(\omega_{A}^{2} + \omega_{B}^{2} - \lambda_{A}^{2} - \lambda_{B}^{2}) - \sqrt{Q}} \label{eq:A24}                        \\
    k_{3} & = \frac{1}{\sqrt{2}}\sqrt{(\omega_{A}^{2} + \omega_{B}^{2} - \lambda_{A}^{2} - \lambda_{B}^{2}) + \sqrt{Q}} \label{eq:A25}                        \\
    Q     & = (\omega_{A}^{2} + \omega_{B}^{2} - \lambda_{A}^{2} - \lambda_{B}^{2})^{2} -                                      \nonumber                      \\
          & 4(\omega_{A} - \lambda_{A})(\omega_{B} - \lambda_{B})[(\omega_{A} + \lambda_{A})(\omega_{B} + \lambda_{B}) - 4(g_{A}^{2} + g_{B}^{2})]  \nonumber
\end{align}
For the symplectic eigenvalues to be real positive, we require \(Q \geq 0\). This imposes the following constraint on the system parameters,
\begin{align}
    g_{A}^{2} + g_{B}^{2} \leq \frac{1}{4}(\omega_{A} + \lambda_{A})(\omega_{B} + \lambda_{B}) \label{eq:A26}
\end{align}
With the symplectic eigenvalues already at hand, we can go ahead to solve the set of Eqs.~\eqref{eq:A16}, however, now we have to eliminate the three redundant equations and only solve for the 9 unknowns from 9 equations.
Now using these equations, we can write down the following equations,
\begin{align}
    \b_{11} & = \frac{\a_{11}(k_{1} - \omega_{A} + \lambda_{A})}{(k_{1}+\omega_{A}-\lambda_{A})} \label{eq:A27} \\
    \b_{12} & = \frac{\a_{12}(k_{1} - \omega_{B} + \lambda_{B})}{(k_{1}+\omega_{B}-\lambda_{B})} \label{eq:A28} \\
    \b_{13} & = \frac{\a_{13}(k_{1} - \omega_{A} + \lambda_{A})}{(k_{1}+\omega_{A}-\lambda_{A})} \label{eq:A29} \\
    \b_{21} & = \frac{\a_{21}(k_{2} - \omega_{A} + \lambda_{A})}{(k_{2}+\omega_{A}-\lambda_{A})} \label{eq:A30} \\
    \b_{22} & = \frac{\a_{22}(k_{2} - \omega_{B} + \lambda_{B})}{(k_{2}+\omega_{B}-\lambda_{B})} \label{eq:A31} \\
    \b_{23} & = \frac{\a_{23}(k_{2} - \omega_{A} + \lambda_{A})}{(k_{2}+\omega_{A}-\lambda_{A})} \label{eq:A32} \\
    \b_{31} & = \frac{\a_{31}(k_{3} - \omega_{A} + \lambda_{A})}{(k_{3}+\omega_{A}-\lambda_{A})} \label{eq:A33} \\
    \b_{32} & = \frac{\a_{32}(k_{3} - \omega_{B} + \lambda_{B})}{(k_{3}+\om_{B}-\lambda_{B})} \label{eq:A34}    \\
    \b_{33} & = \frac{\a_{33}(k_{3} - \omega_{A} + \lambda_{A})}{(k_{3}+\omega_{A}-\lambda_{A})} \label{eq:A35}
\end{align}
Thus, we find the \(\boldsymbol{\b}\) matrix coefficients to be proportional to the corresponding coefficients of the \(\boldsymbol{\a}\) matrix. Now we can use the above in the Bogoliubov identities to eliminate all the \(\boldsymbol{\b}\) coefficients,
\begin{widetext}
    \begin{align}
         & (\a_{11}^{2} + \a_{13}^{2})\frac{4k_{1}(\omega_{A}-\lambda_{A})}{(k_{1}+\omega_{A}-\lambda_{A})^{2}} \;+   \a_{12}^{2}\frac{4k_1(\omega_{B}-\lambda_{B})}{(k_{1}+\omega_{B}-\lambda_{B})^{2}}      = 1 \label{eq:A36} \\
         & (\a_{21}^{2} + \a_{23}^{2})\frac{4k_{2}(\omega_{A}-\lambda_{A})}{(k_{2}+\omega_{A}-\lambda_{A})^{2}} \;+ \a_{22}^{2}\frac{4k_{2}(\omega_{B} - \lambda_{B})}{(k_{2}+\omega_{B}-\lambda_{B})^{2}}  = 1 \label{eq:A37}   \\
         & (\a_{31}^{2} + \a_{33}^{2})\frac{4k_{3}(\omega_{A}-\lambda_{A})}{(k_{3}+\omega_{A}-\lambda_{A})^{2}} \;+ \a_{32}^{2}\frac{4k_{3}(\omega_{B} - \lambda_{B})}{(k_{3}+\omega_{B}-\lambda_{B})^{2}} = 1 \label{eq:A38}
    \end{align}
\end{widetext}
The above equations arise from the diagonal constraints obtained by substituting the \(\boldsymbol{\b}\) matrix coefficients in Eq.~\eqref{eq:A19a}. The off-diagonal constraints give the following:
\begin{widetext}
    \begin{align}
         & \frac{(\a_{11}\a_{21} + \a_{13}\a_{23})(\omega_{A} - \lambda_{A})}{(k_{1}+\omega_{A}-\lambda_{A})(k_{2}+\omega_{A}-\lambda_{A})} \;+ \frac{\a_{12}\a_{22}(\omega_{B}-\lambda_{B})}{(k_{1}+\omega_{B}-\lambda_{B})(k_{2}+\omega_{B}-\lambda_{B})}  = 0 \label{eq:A39}  \\
         & \frac{(\a_{11}\a_{31} + \a_{13}\a_{33})(\omega_{A} - \lambda_{A})}{(k_{1}+\omega_{A}-\lambda_{A})(k_{3}+\omega_{A}-\lambda_{A})} \;+\frac{\a_{12}\a_{32}(\omega_{B}-\lambda_{B})}{(k_{1}+\omega_{B}-\lambda_{B})(k_{3}+\omega_{B}-\lambda_{B})}  = 0 \label{eq:A40}   \\
         & \frac{(\a_{21}\a_{31} + \a_{23}\a_{33})(\omega_{A} - \lambda_{A})}{(k_{2}+\omega_{A}-\lambda_{A})(k_{3}+\omega_{A}-\lambda_{A})}\; + \frac{\a_{22}\a_{32}(\omega_{B}-\lambda_{B})}{(k_{2}+\omega_{B}-\lambda_{B})(k_{3}+\omega_{B}-\lambda_{B})}  =  0 \label{eq:A41}
    \end{align}
\end{widetext}
There are three more equations from the second Bogoliubov identities, however they turn out to be exactly the same as the above equations, as already expected since we obtained the symplectic eigenvalues independent of these equations. Now using the above equations and after a really lengthy algebra, we obtain the following for the \(\a\) and \(\b\) coefficients,

\begin{widetext}
    \begin{center}
        \begin{table}[h!]
            \centering
            \renewcommand{\arraystretch}{2.2}
            \setlength{\tabcolsep}{12pt}
            \caption{\label{table:1}Expressions for the Bogoliubov coefficients \(\alpha_{ij}\) and \(\beta_{ij}\).}
            \begin{tabular}{|c|c|c|c|}
                \hline
                                                                                                    & \(j = 1\) & \(j = 2\) & \(j = 3\) \\
                \hline
                \midrule
                \(\alpha_{1j}\)                                                                     &
                \(\dfrac{(k_1 + \omega_1 - \lambda_1)x_1}{2\sqrt{k_1(\omega_1 - \lambda_1)}}\)      &
                0                                                                                   &
                \(-\dfrac{(k_1 + \omega_1 - \lambda_1)y_1}{2\sqrt{k_1(\omega_1 - \lambda_1)}}\)                                         \\[9pt]
                \hline
                \(\alpha_{2j}\)                                                                     &
                \(\dfrac{(k_2 + \omega_1 - \lambda_1)y_1 y_2}{2\sqrt{k_2(\omega_1 - \lambda_1)}}\)  &
                \(\dfrac{(k_2 + \omega_2 - \lambda_2)x_2}{2\sqrt{k_2(\omega_2 - \lambda_2)}}\)      &
                \(\dfrac{(k_2 + \omega_1 - \lambda_1)x_1 y_2}{2\sqrt{k_2(\omega_1 - \lambda_1)}}\)                                      \\[9pt]
                \hline
                \(\alpha_{3j}\)                                                                     &
                \(-\dfrac{(k_3 + \omega_1 - \lambda_1)y_1 x_2}{2\sqrt{k_3(\omega_1 - \lambda_1)}}\) &
                \(\dfrac{(k_2 + \omega_2 - \lambda_2)y_2}{2\sqrt{k_3(\omega_2 - \lambda_2)}}\)      &
                \(-\dfrac{(k_3 + \omega_1 - \lambda_1)x_1 x_2}{2\sqrt{k_3(\omega_1 - \lambda_1)}}\)                                     \\[9pt]

                \hline

                \(\beta_{1j}\)                                                                      &
                \(\dfrac{(k_1 - \omega_1 + \lambda_1)x_1}{2\sqrt{k_1(\omega_1 - \lambda_1)}}\)      &
                0                                                                                   &
                \(-\dfrac{(k_1 - \omega_1 + \lambda_1)y_1}{2\sqrt{k_1(\omega_1 - \lambda_1)}}\)                                         \\[9pt]
                \hline
                \(\beta_{2j}\)                                                                      &
                \(\dfrac{(k_2 - \omega_1 + \lambda_1)y_1 y_2}{2\sqrt{k_2(\omega_1 - \lambda_1)}}\)  &
                \(\dfrac{(k_2 - \omega_2 + \lambda_2)x_2}{2\sqrt{k_2(\omega_2 - \lambda_2)}}\)      &
                \(\dfrac{(k_2 - \omega_1 + \lambda_1)x_1 y_2}{2\sqrt{k_2(\omega_1 - \lambda_1)}}\)                                      \\[9pt]
                \hline
                \(\beta_{3j}\)                                                                      &
                \(-\dfrac{(k_3 - \omega_1 + \lambda_1)y_1 x_2}{2\sqrt{k_3(\omega_1 - \lambda_1)}}\) &
                \(\dfrac{(k_2 - \omega_2 + \lambda_2)y_2}{2\sqrt{k_3(\omega_2 - \lambda_2)}}\)      &
                \(-\dfrac{(k_3 - \omega_1 + \lambda_1)x_1 x_2}{2\sqrt{k_3(\omega_1 - \lambda_1)}}\)                                     \\[9pt]
                \hline
            \end{tabular}
        \end{table}
    \end{center}
\end{widetext}
Where the phase angles \(\th_{1}\), \(\th_{2}\) are defined as,
\begin{align}
    x_{1} & = \cos{\th_{1}}  = \frac{g_{A}^2}{\sqrt{g_{A}^{2}+g_{B}^{2}}} \label{eq:A42}                                                                  \\
    y_{1} & = \sin{\th_{1}}  = \frac{g_{B}^{2}}{\sqrt{g_{A}^{2} + g_{B}^{2}}}  \label{eq:A43}                                                             \\
    x_{2} & = \cos{\th_{2}}  = \frac{(k_{2}^{2} + \lambda_{A}^{2} - \omega_{A}^{2})}{\sqrt{T}}                                            \label{eq:A44}  \\
    y_{2} & = \sin{\th_{2}}  = \frac{2\sqrt{(\omega_{A} - \lambda_{A})(\omega_{B} - \lambda_{B})(g_{A}^{2} + g_{B}^{2})}}{\sqrt{T}}   \label{eq:A45}      \\
    T     & = (k_{2}^{2} + \lambda_{A}^{2} - \omega_{A}^{2})^{2} + 4(\omega_{A} - \lambda_{A})(\omega_{B} - \lambda_{B})(g_{A}^{2} + g_{B}^{2}) \nonumber
\end{align}
With the above Bogoliubov coefficients and the eigenmode frequencies, we can use the set of Eqs.~\eqref{eq:A15} to solve for the dymanics of the system for any given initial state.
\section{Quantum Measures}
\label{sec:QMeasures}

To quantify quantum correlations between subsystems \textit{A} and \textit{C}, we compute the logarithmic negativity \(\mathcal{E}_{\mathcal{N}}(t)\) and quantum discord \(\mathcal{D}(t)\), both derived from the reduced covariance matrix \(\boldsymbol{\sigma}_{A-C}(t)\), obtained by tracing out the mediator \textit{B}.

For Gaussian states, entanglement is evaluated via the partial transpose operation, implemented by applying the matrix \(\Lambda\)~\cite{Simon:1999lfr}, yielding the partially transposed covariance matrix:
\[
    \tilde{\boldsymbol{\sigma}}_{A-C}(t) = \Lambda\, \boldsymbol{\sigma}_{A-C}(t)\, \Lambda.
\]
The logarithmic negativity is computed from the symplectic eigenvalues \(\tilde{\nu}_i(t)\) of \(\tilde{\boldsymbol{\sigma}}_{A-C}(t)\), which are the eigenvalues of \(i\Omega'\tilde{\boldsymbol{\sigma}}_{A-C}(t)\), where \(\Omega'\) is the reduced symplectic form~\cite{Vidal:2002zz, Plenio2005}. Explicitly,
\begin{align}
    \mathcal{E}_{\mathcal{N}}(t) = \text{max}\left[0,-\sum_{i:\tilde{\nu_i(t)<1}}\log_2(\tilde{\nu}_i(t))\right] \label{eq:A46}
\end{align}

Quantum discord \(\mathcal{D}(t)\) is computed from \(\boldsymbol{\sigma}_{A-C}(t)\) :
\begin{align}
    \mathcal{D}(t) = & f\left(\sqrt{B(t)}\right) - f\left(\nu_{-}(t)\right) - f\left(\nu_{+}(t)\right) \nonumber \\  + &\inf_{\boldsymbol{\sigma}_0} f\left(\sqrt{\det \epsilon}\right)\label{eq:A47}
\end{align}
where \(f(x) = \frac{x+1}{2} \log_2 \left(\frac{x+1}{2}\right) - \frac{x-1}{2} \log_2 \left(\frac{x-1}{2}\right)\), \(B(t)\) is the determinant of the correlation block in \(\boldsymbol{\sigma}_{A-C}(t)\), and \(\nu_{\pm}(t)\) are the symplectic eigenvalues of \(\boldsymbol{\sigma}_{A-C}(t)\), computed as the eigenvalues of \(i\Omega'\boldsymbol{\sigma}_{A-C}(t)\). Note that while the symplectic spectrum of the partially transposed covariance matrix may contain negative entries due to the mathematical diagonalization procedure, the physical symplectic eigenvalues relevant for computing the logarithmic negativity are defined as the absolute values of these quantities.
The final term involves minimizing the determinant of the conditional covariance matrix over all single-mode Gaussian states \(\boldsymbol{\sigma}_0\) and is defined as
\begin{equation}
    \epsilon = \boldsymbol{\sigma}_{A} - \boldsymbol{\sigma}_{B} (\boldsymbol{\sigma}_{C} + \boldsymbol{\sigma}_0)^{-1} \boldsymbol{\sigma}_{B}^{\mathrm{T}}\label{eq:A48}
\end{equation}
where \(\boldsymbol{\sigma}_{A}\), \(\boldsymbol{\sigma}_{B}\), and \(\boldsymbol{\sigma}_{C}\) are the block matrices of the reduced covariance matrix \(\boldsymbol{\sigma}_{A-C}(t)\):
\begin{equation}
    \boldsymbol{\sigma}_{A-C}(t) =
    \begin{pmatrix}
        \boldsymbol{\sigma}_{A}              & \boldsymbol{\sigma}_{B}               \\
        \boldsymbol{\sigma}_{B}^{\mathrm{T}} & \boldsymbol{\sigma}_{C}\label{eq:A49}
    \end{pmatrix},
\end{equation}
and \(\boldsymbol{\sigma}_0\) corresponds to the covariance matrix of a single-mode Gaussian measurement on subsystem \textit{C}. The minimum value of the determinant of \(\epsilon\) is given in Eq.~\eqref{eq:A55}, where \(\mathcal{A} = \det(\boldsymbol{\sigma}_{A})\), \(\mathcal{B} = \det(\boldsymbol{\sigma}_{B})\), \(\C = \det(\boldsymbol{\sigma}_{C})\), and \(\D = \det(\boldsymbol{\sigma}_{A-C})\).
The fidelity between two covariance matrices \(V_1\) and \(V_2\) is defined as~\cite{Banchi:2015}:
\begin{align}
    F_0(V_1, V_2)    & = \frac{F_{\text{tot}}}{\sqrt[4]{\det(V_1 + V_2)}}\label{eq:A50}                                                                               \\
    F_{\text{tot}}^4 & = \det \left[ 2 \left( \sqrt{ \mathbb{I} + \frac{(V_{\text{aux}} \Omega)^{-2}}{4} } + \mathbb{I} \right) V_{\text{aux}} \right] \label{eq:A51} \\
                     & = \det \left[ \left( \sqrt{ \mathbb{I} - W_{\text{aux}}^{-2} } + \mathbb{I} \right) W_{\text{aux}}\, i\Omega \right] \label{eq:A52}
\end{align}
where the auxiliary matrices \(V_{\text{aux}}\) and \(W_{\text{aux}}\) are given by
\begin{align}
    V_{\text{aux}} & := \Omega^{T} (V_1 + V_2)^{-1} \left( \frac{\Omega}{4} + V_2 \Omega V_1 \right)\label{eq:A53} \\
    W_{\text{aux}} & := -2 V_{\text{aux}} i \Omega = -(W_1 + W_2)^{-1} ( \mathbb{I} + W_2 W_1 )\label{eq:A54}
\end{align}
Here, \(\Omega\) is the symplectic form, and \(W_i := -2 V_i i \Omega\).
\begin{widetext}
    \begin{equation}
        E^{\text{min}} = \inf_{\boldsymbol{\sigma}_0} \det(\epsilon) =
        \begin{cases}
            \displaystyle \frac{2\C^2 + (-1 + \B)(-\A + \D) + 2|\C| \sqrt{\C^2 + (-1 + \B)(-\A + \D)}}{(-1 + \B)^2} & \text{if } (\D - \A\B)^2 \leq (1 + \B)\C^2(\A + \D) \\
            \displaystyle \frac{\A\B - \C^2 + \D - \sqrt{\C^4 + (-\A\B + \D)^2 - 2\C^2(\A\B + \D)}}{2\B}            & \text{otherwise}
        \end{cases}
        \label{eq:A55}
    \end{equation}
\end{widetext}

\section{Log Negativity in the weak coupling limit}
\label{sec:LogNeg}

In this section, we provide the details on the calculation of \(\langle\mathcal{E}_{\mathcal{N}}\rangle\) in the weak-coupling regime with symmetric oscillators with \(\om_A = \om_B = \om\), \(\la_A = \la_B = 0\), \(g_A = g_B = g\) and we define the dimensionless coupling parameter \(\la_p = \sqrt{2}g/\om\). In this limit, we perform a perturbative expansion of the eigenmode frequencies \(k_{1}, k_{2}, k_{3}\) [Eqs.~\eqref{eq:A23}--\eqref{eq:A25}] and the Bogoliubov coefficients (Table~\ref{table:1}) in powers of \(\lambda_p\), retaining only terms up to linear order and discarding higher-order corrections.
\begin{align}
    \boldsymbol{\sigma}_{A-C}(t) \approx \mathbb{I}_4 + \la_p\Delta + \mathcal{O}({\la_p^2})\label{eq:A56}
\end{align}
where \(\mathbb{I}_4\) is \(4\times4\) identity matrix and \(\Delta\) is defined as,
\begin{align}
    \Delta = -\sin{(2\om \la_pt)}R\begin{pmatrix}
                                      \mathbb{I}_2 & \mathbb{I}_2 \\
                                      \mathbb{I}_2 & \mathbb{I}_2
                                  \end{pmatrix}\nonumber \\
    R = \begin{pmatrix}
            \sin{(2\om t)} & \cos{(2\om t)}  \\
            \cos{(2\om t)} & -\sin{(2\om t)}
        \end{pmatrix} \label{eq:A57}
\end{align}
with \(\mathbb{I}_2\) being the \(2\times2\) identity matrix. The symplectic eigenvalues of the partial transposed reduced covariance matrix \(\tilde{\boldsymbol{\sigma}}_{A-C}(t)\) then becomes,
\begin{subequations}
    \label{eq:A58}
    \begin{eqnarray}
        \tilde{\nu}_{1+} &\approx& \sqrt{1 + \la_p\sin{(2\om \la_p t)}} \label{eq:A58a}\\
        \tilde{\nu}_{1-} &\approx& -\sqrt{1 + \la_p\sin{(2\om \la_p t)}} \label{eq:A58b}\\
        \tilde{\nu}_{2+} &\approx& \sqrt{1 - \la_p\sin{(2\om \la_p t)}} \label{eq:A58c}\\
        \tilde{\nu}_{2-} &\approx& -\sqrt{1 - \la_p\sin{(2\om \la_p t)}} \label{eq:A58d}
    \end{eqnarray}
\end{subequations}
Using Eqs.~\eqref{eq:A58} in Eq.~\eqref{eq:A46}, we obtain,
\begin{align}
    \mathcal{E}_{\mathcal{N}}(t) \approx \frac{\lambda_{p}|\sin{(2\sqrt{2}g t)}|} {2\ln{(2)}}  + \mathcal{O}({\la_p^2})\label{eq:A59}
\end{align}
Thus, the average value of \(\mathcal{E}_{\mathcal{N}}(t)\) in the weak-coupling regime becomes,
\begin{align}
    \langle \mathcal{E}_{\mathcal{N}} \rangle = \frac{\lambda_p}{\pi \ln 2} =\frac{\sqrt{2}}{\pi \ln 2}\frac{g}{\omega}\label{eq:A60}
\end{align}
\begin{figure*}[!htbp]
    \centering
    \begin{subfigure}[t]{0.24\textwidth}
        \centering
        \includegraphics[width=\textwidth]{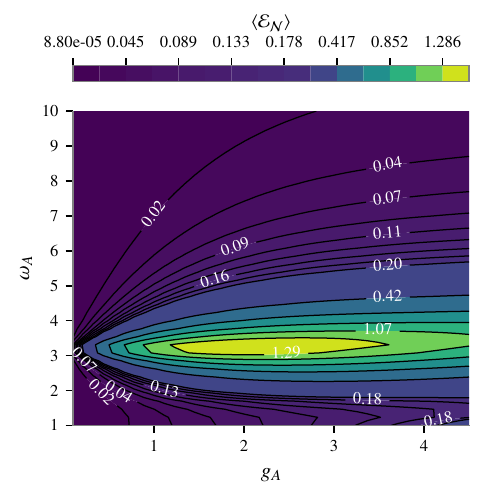}
        \caption{}
    \end{subfigure}
    \begin{subfigure}[t]{0.24\textwidth}
        \centering
        \includegraphics[width=\textwidth]{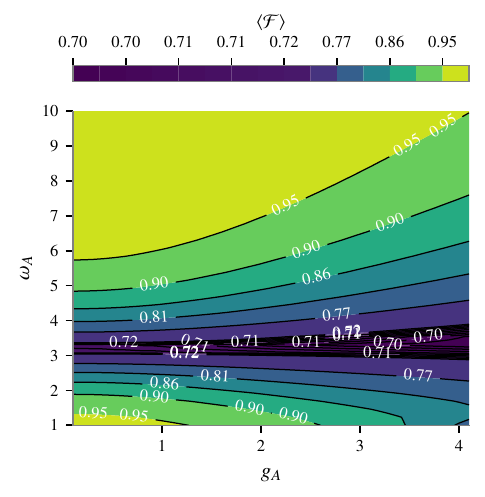}
        \caption{}
    \end{subfigure}
    \begin{subfigure}[t]{0.24\textwidth}
        \centering
        \includegraphics[width=\textwidth]{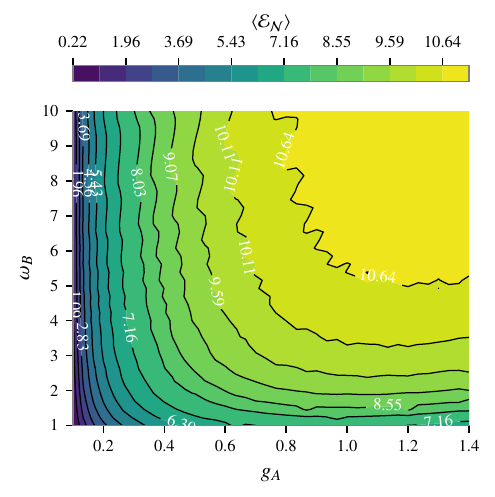}
        \caption{}
    \end{subfigure}
    \begin{subfigure}[t]{0.24\textwidth}
        \centering
        \includegraphics[width=\textwidth]{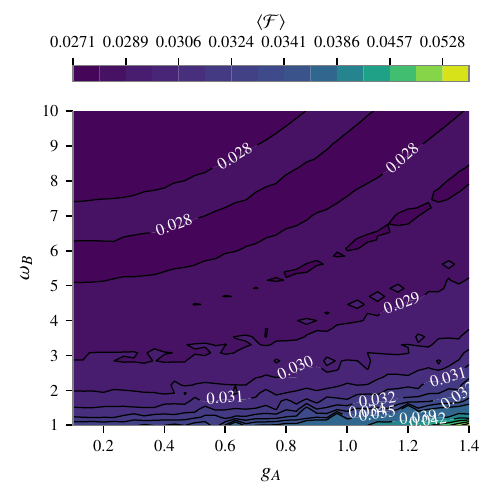}
        \caption{}
    \end{subfigure}
    \caption{\label{fig:4} Contour plots of time-averaged logarithmic negativity \(\langle \mathcal{E}_{\mathcal{N}} \rangle\) and Fidelity \(\langle \mathcal{F}\rangle\).
    \textbf{(a, b)}(HMR) use the parameter set \(\omega_B = 50\), \(\lambda_B = 49.9\), \(\lambda_A = 0.001\), \(g_B = 2.0\),
    \textbf{(c, d)}(LMR) use \(\omega_A = 5.0\), \(\lambda_A = 4.999\), \(\lambda_B = \omega_B/1.5\), \(g_B = 1.4\)
    }
\end{figure*}
\begin{figure*}[!htbp]
    \centering
    \begin{subfigure}[t]{0.24\textwidth}
        \centering
        \includegraphics[width=\textwidth]{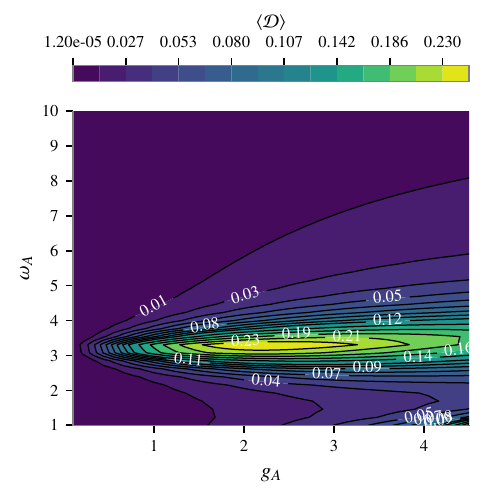}
        \caption{\scriptsize HMR (\(\omega_B = 50.0\), \(\lambda_B = 49.9\), \(\lambda_A = 0.001\), \(g_B = 2.0\))}
        \label{Fig3:HMRa}
    \end{subfigure}
    \begin{subfigure}[t]{0.24\textwidth}
        \centering
        \includegraphics[width=\textwidth]{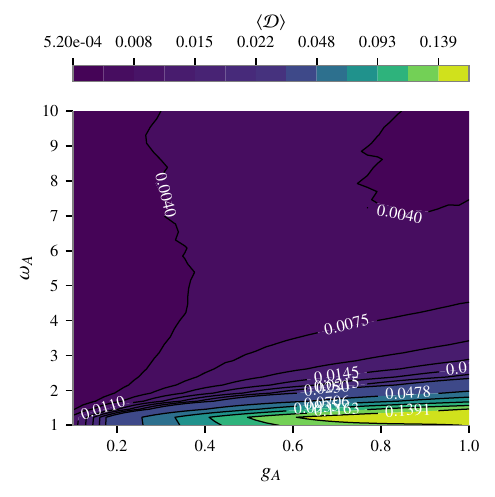}
        \caption{\scriptsize HMR (\(\omega_B = 5.0\), \(\lambda_B = 4.9\), \(\lambda_A = 0.0001\), \(g_B = 1.0\))}
        \label{Fig3:HMRb}
    \end{subfigure}
    \begin{subfigure}[t]{0.24\textwidth}
        \centering
        \includegraphics[width=\textwidth]{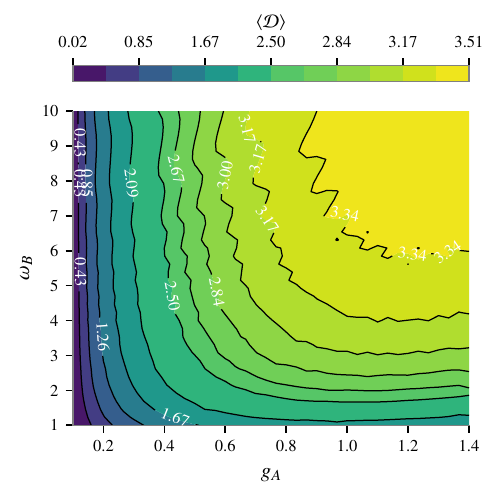}
        \caption{\scriptsize LMR (\(\omega_A = 5.0\), \(\lambda_A = 4.999\), \(\lambda_B = \omega_B/1.5\), \(g_B = 1.4\))}
        \label{Fig3:LMRa}
    \end{subfigure}
    \begin{subfigure}[t]{0.24\textwidth}
        \centering
        \includegraphics[width=\textwidth]{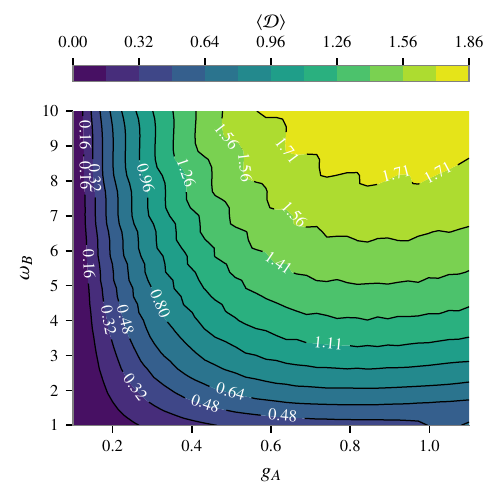}
        \caption{\scriptsize LMR (\(\omega_A = 5.0\), \(\lambda_A = 4.999\), \(\lambda_B = \omega_B/1.5\), \(g_B = 1.4\))}
        \label{Fig3:LMRb}
    \end{subfigure}
    \caption{\label{fig:5}Contour plots of time-averaged quantum discord \(\langle \mathcal{D} \rangle\) in two contrasting mediator regimes.
        \textbf{(a, c)} Vacuum initial state, corresponding to zero squeezing: \(r_1 = r_2 = r_3 = 0\), \(\theta_1 = \theta_2 = \theta_3 = 0\);
        \textbf{(b, d)} Squeezed initial state with \(r_1 = (0.1, 0.5)\), \(r_2 = (1.0, 2.0)\), \(r_3 = (0.1, 0.5)\) and squeezing angles \(\theta_1 = (-\pi/3, 0)\), \(\theta_2 = (\pi/4, 0)\), \(\theta_3 = (\pi/6, 0)\).}
\end{figure*}
\section{Plots for Vacuum Initial state}
\label{sec:Plots}

In Fig.~\eqref{fig:4} we provide the contour maps for \(\langle \mathcal{E}_{\mathcal{N}}\rangle\), and \(\langle \mathcal{F}\rangle\) for the case where the initial state of the system is vacuum. For the case of initial squeezed states, the plots were provided in Fig.~\ref{fig:2}. In Fig.~\ref{fig:5} above we provide the countour plots for quantum discord \(\langle \mathcal{D}\rangle\) both for initial squeezed state and initial vacuum state. Note that the behaviour of quantum discord resembles that of logarithmic negativity.

\section{Exact Solution for the HMR Limit}
\label{sec:HMR}
In this section, we present the exact solution for the heavy mediator regime (HMR), characterized by the condition \(\omega_B = \lambda_B\). In this limit, the eigenmode frequencies simplify to \(k_1 = k_3 = \sqrt{\omega_A^2 - \lambda_A^2}\) and \(k_2 = 0\), with the vanishing of \(k_2\) indicating a frozen mode. Following the same procedure outlined in Appendix.~\ref{sec:3CHOsol}, and substituting \(\omega_B = \lambda_B\), the relations between the Bogoliubov coefficients \(\alpha_{ij}\) and \(\beta_{ij}\) reduce to:
\begin{align}
    \beta_{11} & = \frac{\alpha_{11}(k_{1} - \omega_{A} + \lambda_{A})}{(k_{1}+\omega_{A}-\lambda_{A})} \label{eq:A61} \\
    \beta_{12} & = \alpha_{12} \label{eq:A62}                                                                          \\
    \beta_{13} & = \frac{\alpha_{13}(k_{1} - \omega_{A} + \lambda_{A})}{(k_{1}+\omega_{A}-\lambda_{A})} \label{eq:A63} \\
    \beta_{21} & = -\alpha_{21} \label{eq:A64}                                                                         \\
    \beta_{23} & = -\alpha_{23} \label{eq:A65}                                                                         \\
    \beta_{31} & = \frac{\alpha_{31}(k_{3} - \omega_{A} + \lambda_{A})}{(k_{3}+\omega_{A}-\lambda_{A})} \label{eq:A66} \\
    \beta_{32} & = \alpha_{32} \label{eq:A67}                                                                          \\
    \beta_{33} & = \frac{\alpha_{33}(k_{3} - \omega_{A} + \lambda_{A})}{(k_{3}+\omega_{A}-\lambda_{A})} \label{eq:A68}
\end{align}
While this method does not yield a direct relation for \(\beta_{22}\), we may introduce the parametrization \(\beta_{22} = \chi_2 \alpha_{22}\) with \(|\chi_2| \geq 1\) without loss of generality. The full set of Bogoliubov coefficients is summarized in Table~\ref{table:2}, where \(\chi_1\) is an arbitrary real parameter.
Interestingly, in this limit the only constraint arising from the positivity of the eigenfrequencies is \(\omega_A \ge \lambda_A\). This permits arbitrarily large values for the subsystem-mediator couplings \(g_A\) and \(g_B\). However, despite this strong coupling, the structure of the Bogoliubov transformation in this regime prohibits any entanglement transfer between the subsystems, effectively rendering the mediator dynamically inert. This can be verified analytically by computing the symplectic eigenvalues of the partial transposed reduced covariance matrix, which turn out to be
\begin{align*}
    \tilde{\nu}_1 & = 1                                                                                                                                                             \\
    \tilde{\nu}_2 & = \sqrt{1 + \frac{16 \sin^2\left( \tfrac{1}{2}k_1 t \right)(\cos{(k_1 t)}\lambda_A + \omega_A)(g_A^2 + g_B^2)}{(\omega_A - \lambda_A)(\omega_A + \lambda_A)^2}}
\end{align*}
Since \(\omega_A > \lambda_A\), both symplectic eigenvalues are always positive and greater than 1.

\begin{figure}[t]
    \centering
    \includegraphics[width=0.49\linewidth]{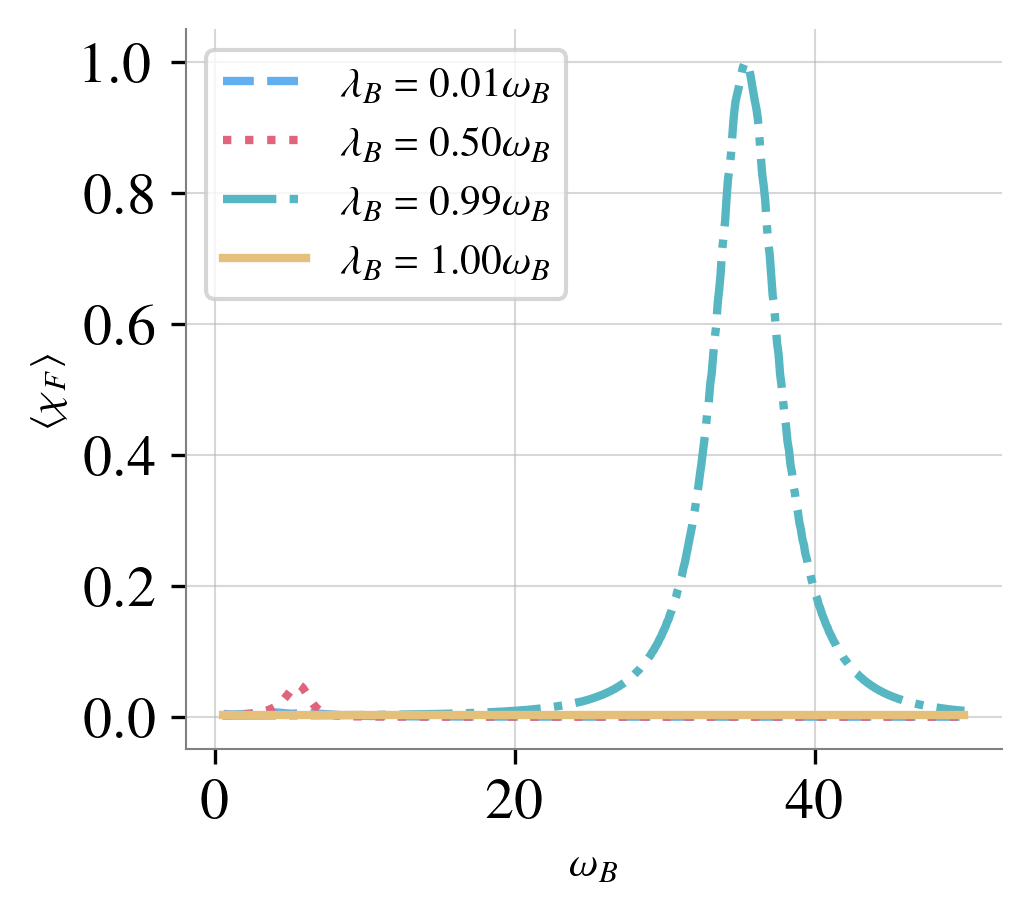}
    \includegraphics[width=0.49\linewidth]{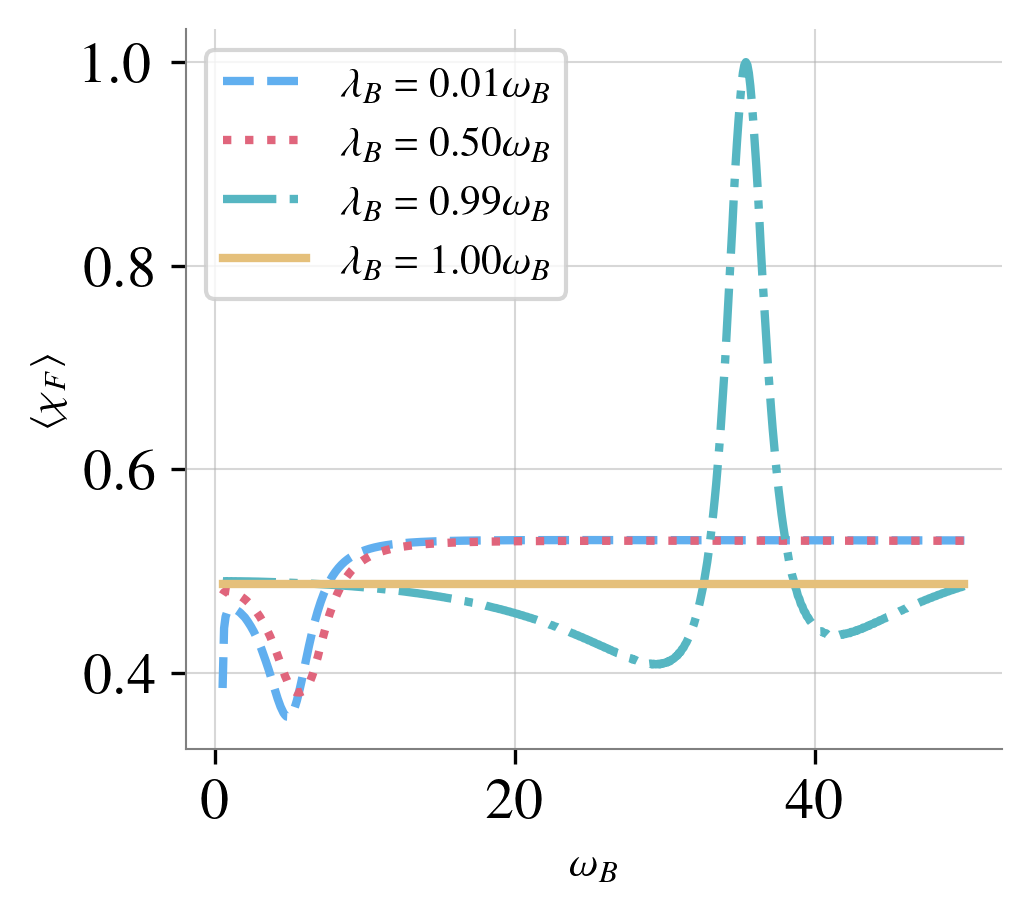}
    \caption{\label{fig:6}
        Normalized time-averaged dynamical fidelity susceptibility vs system parameters.
        Left: vacuum initial state; right: squeezed initial state with $r_1 = r_3 = 0.5$, $r_2 = 0.2$, \(\th_1 = \th_2 = \th_3 = 0\).
        Fixed parameters: $\omega_A = 5$, $\lambda_A = 0.01\,\omega_A$, $g_A = g_B = 0.5$.}
\end{figure}
\section{Dynamical Fidelity Susceptibility}
\label{sec:DSF}
The time-averaged dynamical fidelity susceptibility $\langle \chi_F \rangle$ provides a potential method for identifying the mediator's regime in BMV-type experiments. In this section, we present the results for the variation of $\langle \chi_F \rangle$ with $\omega_B$ in the case $\lambda_A = 0.01\,\omega_A$. As shown in Fig.~\ref{fig:6}, the qualitative features closely resemble those observed for $\lambda_A = 0.5\,\omega_A$. In particular, the stability of $\langle \chi_F \rangle$ in the HMR limit under variations in $\omega_B$ near $\omega_A$ suggests that this behavior can serve as a reliable diagnostic for characterizing the nature of the mediator.

\begin{widetext}
    \begin{center}
        \begin{table}[h!]
            \centering
            \renewcommand{\arraystretch}{2.2}
            \setlength{\tabcolsep}{12pt}
            \caption{\label{table:2}Expressions for the Bogoliubov coefficients \(\alpha_{ij}\) and \(\beta_{ij}\) in the HMR limit.}
            \begin{tabular}{|c|c|c|c|}
                \hline
                                                                                                                      & \(j = 1\) & \(j = 2\) & \(j = 3\) \\
                \hline
                \midrule
                \(\alpha_{1j}\)                                                                                       &
                \(\dfrac{\chi_1(k_1 + \omega_A - \lambda_A)}{2\sqrt{1 + \chi_1^2}\sqrt{k_1(\omega_A - \lambda_A)}}\)  &
                \(\dfrac{k_1(g_B + \chi_1g_A)}{(\om_A + \la_A )\sqrt{1 + \chi_1^2}\sqrt{k_1(\om_A - \la_A)}}\)        &
                \(\dfrac{(k_1 + \omega_A - \lambda_A)}{2\sqrt{1 +  \chi_1^2}\sqrt{k_1(\omega_1 - \lambda_1)}}\)                                           \\[9pt]
                \hline
                \(\alpha_{2j}\)                                                                                       &
                \(\dfrac{-(\chi_2 -1)g_A}{(\om
                _A + \la_A)\sqrt{\chi_2^2 - 1}}\)                                                                     &
                \(\dfrac{\chi_2}{\sqrt{\chi_2^2 - 1}}\)                                                               &
                \(\dfrac{-(\chi_2 -1)g_B}{(\om
                _A + \la_A)\sqrt{\chi_2^2 - 1}}\)                                                                                                         \\[9pt]
                \hline
                \(\alpha_{3j}\)                                                                                       &
                \(\dfrac{(k_1 + \omega_A - \lambda_A)}{2\sqrt{1 + \chi_1^2}\sqrt{k_1(\omega_1 - \lambda_1)}}\)        &
                \(\dfrac{k_1(g_A - \chi_1g_B)}{2(\om_A + \la_A)\sqrt{1 + \chi_1^2}\sqrt{k_1(\omega_A - \lambda_A)}}\) &
                \(-\dfrac{(k_1 + \omega_A - \lambda_A)\chi_1}{2\sqrt{
                1 + \chi_1^2}\sqrt{k_1(\omega_A - \lambda_A)}}\)                                                                                          \\[9pt]

                \hline

                \(\beta_{1j}\)                                                                                        &
                \(\dfrac{\chi_1(k_1 - \omega_A + \lambda_A)}{2\sqrt{1 + \chi_1^2}\sqrt{k_1(\omega_A - \lambda_A)}}\)  &
                \(\dfrac{k_1(g_B + \chi_1g_A)}{(\om_A + \la_A )\sqrt{1 + \chi_1^2}\sqrt{k_1(\om_A - \la_A)}}\)        &
                \(\dfrac{(k_1 - \omega_A + \lambda_A)}{2\sqrt{1 +  \chi_1^2}\sqrt{k_1(\omega_1 - \lambda_1)}}\)                                           \\[9pt]
                \hline
                \(\beta_{2j}\)                                                                                        &
                \(\dfrac{(\chi_2 -1)g_A}{(\om
                _A + \la_A)\sqrt{\chi_2^2 - 1}}\)                                                                     &
                \(\dfrac{1}{\sqrt{\chi_2^2 -1 }}\)                                                                    &
                \(\dfrac{(\chi_2 -1)g_B}{(\om
                _A + \la_A)\sqrt{\chi_2^2 - 1}}\)                                                                                                         \\[9pt]
                \hline
                \(\beta_{3j}\)                                                                                        &
                \(\dfrac{(k_1 - \omega_A + \lambda_A)}{2\sqrt{1 + \chi_1^2}\sqrt{k_1(\omega_1 - \lambda_1)}}\)        &
                \(\dfrac{k_1(g_A - \chi_1g_B)}{2(\om_A + \la_A)\sqrt{1 + \chi_1^2}\sqrt{k_1(\omega_A - \lambda_A)}}\) &
                \(-\dfrac{(k_1 - \omega_A + \lambda_A)\chi_1}{2\sqrt{
                1 + \chi_1^2}\sqrt{k_1(\omega_A - \lambda_A)}}\)                                                                                          \\[9pt]
                \hline
            \end{tabular}
        \end{table}
    \end{center}
\end{widetext}

%

\end{document}